\documentclass{article}

\usepackage{arxiv}

\usepackage[utf8]{inputenc} 
\usepackage[T1]{fontenc}    
\usepackage{hyperref}       
\usepackage{url}            
\usepackage{booktabs}       
\usepackage{amsfonts}       
\usepackage{nicefrac}       
\usepackage{microtype}      
\usepackage{lipsum}         
\usepackage{graphicx}
\usepackage{natbib}
\usepackage{hyperref}
    
\usepackage{doi}

\usepackage{amsmath,amssymb}
\usepackage{cleveref}       

\usepackage{appendix} 

\newcommand{\rectangle}{{
  \ooalign{$\sqsubset\mkern3mu$\cr$\mkern3mu\sqsupset$\cr}%
}}

\newcommand{\authorHref}[3][black]{\href{#2}{\color{#1}{#3}}}
\title{Monitoring fluid saturation in reservoirs using time-lapse full-waveform
inversion}

\author{\authorHref{https://orcid.org/0000-0002-0417-1259}{\includegraphics[scale=0.06]{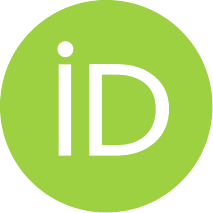}\hspace{1mm}Amir Mardan}\\
	INRS-ETE\\
	Geostack\\
Québec, QC, Canada \\
	\texttt{mardan.amir.h@gmail.com} \\
	\And
\authorHref{https://orcid.org/0000-0002-2042-2759}{\includegraphics[scale=0.06]{orcid.pdf}\hspace{1mm}Bernard Giroux} \\
	INRS-ETE\\
	Québec, QC, Canada \\
	\texttt{bernard.giroux@inrs.ca} \\
	\And
	\authorHref{https://orcid.org/0000-0002-1849-3718}{\includegraphics[scale=0.06]{orcid.pdf}\hspace{1mm}Gabriel Fabien-Ouellet} \\
	Polytechnique Montréal\\
	Montréal, QC, Canada \\
	\texttt{gabriel.fabien-ouellet@polymtl.ca} \\
\And
	\authorHref{https://orcid.org/0000-0002-1849-3718}{\includegraphics[scale=0.06]{orcid.pdf}\hspace{1mm}Mohammad Reza Saberi} \\
	GeoSoftware\\
	The Hague, Netherlands \\
	\texttt{reza.saberi@geosoftware.com}
}

\renewcommand{\shorttitle}{Monitoring fluid saturation in reservoirs using TL-FWI}

\hypersetup{
colorlinks   = true,
linkcolor    = blue,
urlcolor = blue,
citecolor = blue,
pdftitle={Monitoring fluid saturation in reservoirs using time-lapse full-waveform
inversion},
pdfsubject={physics.geo-ph},
pdfauthor={Amir Mardan},
pdfkeywords={full-waveform inversion,
CO\textsubscript{2~}monitoring, time-lapse inversion, reservoir monitoring,
rock-physics monitoring},
}

\graphicspath{ {./figures/} }

\begin{document}

\maketitle

\begin{abstract}
Monitoring the rock-physics properties of the subsurface is of great importance for reservoir management. 
For either oil and gas applications or CO$_2$ storage, seismic data are a valuable source of information for tracking changes in elastic properties which can be related to fluids saturation and pressure changes within the reservoir.
Changes in elastic properties can be estimated with time-lapse full-waveform inversion.
Monitoring rock-physics properties, such as saturation, with time-lapse full-waveform inversion is usually a two-step process: first, elastic properties are estimated with full-waveform inversion, then, the rock-physics properties are estimated with rock-physics inversion. 
However, multiparameter time-lapse full-waveform inversion is prone to crosstalk between parameter classes across different vintages. 
This leads to leakage from one parameter class to another, which, in turn, can introduce large errors in the estimated rock-physics parameters. 
To avoid inaccuracies caused by crosstalk and the two-step inversion strategy, we reformulate time-lapse full-waveform inversion to estimate directly the changes in the rock-physics properties.
Using Gassmann's model, we adopt a new parameterization containing  porosity, clay content, and water saturation.
In the context of reservoir monitoring, changes are assumed to be induced by fluid substitution only.
The porosity and clay content can thus be kept constant during time-lapse inversion. 
We compare this parameterization with the usual density-velocity parameterization for different benchmark models. 
Results indicate that the proposed parameterization eliminates crosstalk between parameters of different vintages, leading to more accurate estimation of saturation changes. 
We also show that using the parameterization based on porosity, clay content, and water saturation,  the elastic changes can be monitored more accurately. 
\end{abstract}



\section*{Introduction}
Either for oil and gas applications or CO$_2$ sequestration, the ability to monitor reservoirs is highly important as it allows for safe and profitable operations.
Time-lapse seismic monitoring is a powerful tool that allows to investigate large-scale changes in a reservoir \cite[]{Lumley2001, LandroEtAl2003,DupuyEtAl2016_2, MaharramovEtAl2016, Fabien_OuelletEtAl2017, lang2019rock, ZhouEtLumley2021,MardanEtAl2022sat_eage, MardanEtAl2022sat_seg, MardanEtAl2022nr_eage}.
Seismic monitoring is sensitive to changes in elastic parameters, which can be related to reservoir parameters, such as saturation, through a rock-physics model.

Numerous studies have been conducted to monitor saturation changes in reservoirs.  
\cite{LandroEtAl2003} used $PP$ and $PS$ seismic data to better estimate the time-lapse changes.  
\cite{BulandAndOuair2006} proposed a Bayesian time-lapse inversion method and applied it to monitor the Norne Field offshore Norway with seismic data acquired from 2001 to 2003.
\cite{VedantiAndSend2009} used prestack seismic data inversion to monitor the in situ combustion in the Balol oil field. 
\cite{lang2019rock} studied the potential of the Johansen field in Norway
by simulating $\text{CO}_2$ injection for 10 years and monitoring the reservoir.
All of these studies are based on linearized amplitude variation with offset (AVO) and the convolutional model whose accuracy decreases by increasing the incidence angle \cite[]{mallick2007amplitude}.
Although AVO has been very successful and is commonly used in industry, it has some other limitations.
AVO inversion only uses the amplitude of the reflected waves while for better imaging, all type of waves should be involved during the inversion \cite[]{VirieuxEtAl2017}.
In addition, AVO relies on the assumption that reflectors are flat, which is a restrictive hypothesis in many cases \cite[]{QueiberEtSingh2013}.
Last but not least, AVO suffers from the uncertainties in data preprocessing such as velocity model errors \cite[]{naeini2017full, HuEtAl2021}.
These assumptions and the incomplete use of waveforms can lead to suboptimal use of the resources invested in monitoring.

A variety of studies have used time-lapse full-waveform inversion (TL-FWI) to monitor reservoirs.   
For example, \cite{WatanabeEtAl2005} used time-lapse crosswell seismic data to monitor the changes during a gas hydrate thermal production test at the Mallik site.
\cite{RaknesEtAl2013} monitored the leakage of one producing well over a field in North Sea by studying the seismic data of two vintages with two years time difference.
\cite{HicksEtAl2016} and \cite{LescoffitEtAl2016} showed the interest of TL-FWI when studying seismic data and changes in $P$-wave velocity ($V_P$) at Grane field due to gas-oil replacement.
\cite{MaharramovEtAl2016} showed the potential of TL-FWI to image the changes in the Genesis field in the Green Canyon area of the central Gulf of Mexico.
In these studies, TL-FWI is carried out to map the acoustic changes in the subsurface using a monoparameter acoustic formulation (analyzing only $V_P$).
To obtain the changes in terms of rock-physics properties, full-waveform inversion (FWI) is employed to estimate $V_P$ and in the next step, the estimated $V_P$ is inverted using rock-physics inversion.
For example, this two-step procedure has been used to estimate the changes in the saturation of CO$_2$ at Sleipner field in North Sea \cite[]{QueiberEtSingh2013, DupuyEtAl2016_2, DupuyEtAl2016_1, yan2019co2,dupuy2021combined, romdhane2022toward}.

To approach the true properties of the subsurface, an elastic formulation is preferred over acoustic as it allows taking into account the different seismic phases in the measured data.
An elastic medium requires multiparameter TL-FWI.
In this case, model parameters can have coupled effects on the seismic data, which is referred to as crosstalk between parameters \cite[]{OpertoEtAl2013, YangEtAl2018}.
\cite{MaEtAl2016} studied the Marmousi model for multiparameter acoustic TL-FWI and showed that the crosstalk problem is more severe in TL-FWI.
Although \cite{MaEtAl2016} recovered good images of the subsurface using FWI, the estimated \textit{time-lapse} images were inaccurate due to the crosstalk problem (Figure 2 and 3 of \cite{MaEtAl2016}).
This problem can be alleviated by using different parameterizations \cite[]{tarantola1986, OpertoEtAl2013, MaEtAl2016,HuEtAl2021}.
A comprehensive discussion is provide by \cite{pan2019interparameter} where the authors analyzed the efficiency of different parameterizations such as velocity-density (DV) ($P$-wave velocity, $S$-wave velocity, and density), modulus-density (DM) (bulk modulus, shear modulus, and density), impedance-density (D-IP) ($P$-wave impedance, $S$-wave impedance, and density),  velocity-impedance-$I$ (V-IP-$I$) ($P$-wave velocity, $S$-wave velocity,
and $P$-wave impedance), and velocity-impedance-$II$ (V-IP-$II$)
($P$-wave velocity, $S$-wave velocity, and $S$-wave impedance) for performing elastic FWI.

The goal of this study is to recover the changes in water saturation ($S_w$) by considering an elastic earth.
To minimize the effects of crosstalk that can occur due to the time-lapse nature of the problem, the TL-FWI is formulated in terms of porosity ($\phi$), clay content ($C$), and water saturation, as proposed by \cite{HuEtAl2021} in the context of FWI.
This parameterization (called PCS) is chosen to facilitate the time-lapse inversion as we can assume that porosity and clay content are constant with time in the reservoir. 
To perform this technique, a connection between the elastic and rock-physical properties is required. 
 In this work, we use Gassmann's equation for this purpose. 
Using the PCS parameterization, we combine the rock-physics inversion and FWI in a single inversion scheme.
In this way, saturation can be directly inverted from the raw seismic data (shot gather).
By performing TL-FWI with the PCS parameterization, we can also easily obtain the time-lapse images of elastic properties using the underlying rock-physics model.

This paper is organized by first presenting FWI and the required formulation to include Gassmann's model for estimating the porosity, clay content, and water saturation.
This section is followed by a brief discussion on TL-FWI.
Then, we study and discuss the efficiency of this method by using numerical models.

\section*{Theory}

\subsection*{Full-waveform inversion}
Full-waveform inversion is a local optimization process that minimizes residuals between the observed and estimated wavefields at the receivers locations for different source positions \cite[]{VirieuxEtOperto2009}.
The objective function can be written as
\begin{equation}
\chi(\mathbf{m}) = \frac{1}{2}\| \mathbf{W}_d\left(\mathbf{R}\mathbf{u}(\mathbf{m}) - \mathbf{d} \right)\|_2^2 + \chi_{REG},\
\label{eq:cost_function}
\end{equation}
where $\mathbf{R}$ maps the wavefield ($\mathbf{u}$) to the receivers locations. 
Vectors $\mathbf{m}$ and $\mathbf{d}$ are the model parameter and observed data, respectively.
$\mathbf{W}_d$ is a weighting operator on the data misfit and $\chi_{REG}$ is a regularization term.
FWI is highly ill-posed, where an infinite number of models matches the data. 
A variety of regularization methods such as the Tikhonov, total variation, prior information, and parameter-relation techniques have been proposed to condition the FWI problem and reduce the ill-posedness \cite[]{VirieuxEtOperto2009,AsnaashariEtAl2015}.
In this study,  we use the Tikhonov regularization.
This regularization method is presented in Appendix \ref{app:regularization} in addition to total variation, prior information, and parameters-relation techniques.
In equation \ref{eq:cost_function}, the wavefield is the solution of a partial differential equation (PDE) which is discretized to perform  the forward modeling.
For the 2D isotropic elastic case in time domain,  this PDE can be written as
\begin{equation}
\begin{cases}
\rho \dot v_x = \partial_x \tau_{xx}+ \partial_z \tau_{xz},\\
\\
\rho \dot v_z = \partial_z \tau_{zz}+ \partial_x \tau_{xz},\\
\\
\dot\tau_{xx}= (\lambda + 2 G) \partial_x v_x+ \lambda \partial_z v_z + s,\\
\\
\dot\tau_{zz}= \lambda \partial_x v_x+ (\lambda + 2 G) \partial_z v_z + s,\\
\\
\dot \tau_{xz} = G (\partial_z v_x + \partial_x v_z),\\
\end{cases}
\label{eq:forward_modeling}
\end{equation}
where $\lambda$ and $G$ denote lamé's parameter and where $\rho$ and $s$ are density and the source function.
In equation \ref{eq:forward_modeling}, the particle velocities in $x$- and $z$-directions ($v_x$ and $v_z$) as well as ($\tau_{xx}$ and $\tau_{zz}$) and shear stresses ($\tau_{xz}$) constitute vector $\mathbf{u}$. 
Spatial partial derivative operators are indicated by $\partial_x$ and $\partial_z$ for $x-$ and $z-$directions and the temporal differentiation is denoted by overhead dot ($\dot \rectangle$).

In the case of equation \ref{eq:forward_modeling}, the vector of model parameter, $\mathbf{m}$ in equation \ref{eq:cost_function}, is
\begin{equation}
\mathbf{m} = [\lambda, G, \rho]^T,
\label{eq:model_parameters_dv}
\end{equation}
where $^T$ is the transpose operator. 
Although,  the forward modeling is performed using Lamé parameters and density, FWI can be performed using different parameterizations.
For example, the DV parameterization is a common formulation.
Switching between parameters can be carried out using the chain rule, as shown at the end of this section.

\subsubsection*{Optimization}
To minimize the cost function (equation \ref{eq:cost_function}), a variety of optimization algorithms such as steepest-descent, conjugate-gradient, $\ell$-BFGS, and Newton algorithms can be employed \cite[]{VirieuxEtOperto2009}.
These algorithms minimize the cost function in the vicinity of a starting model, $\mathbf{m}_0$, as 
\begin{equation}
\mathbf{m} = \mathbf{m}_0 + \alpha \pmb{\Delta} \mathbf{m},
\label{eq:perturbation}
\end{equation}
where $\alpha$ is step size and the search direction, $\pmb{\Delta} \mathbf{m}$, for gradient descent is
\begin{equation}
\pmb{\Delta} \mathbf{m} = -\left[ \begin{array}{c}
\nabla_{\mathbf{m}^{(1)}}\chi\\
\nabla_{\mathbf{m}^{(2)}}\chi\\
\nabla_{\mathbf{m}^{(3)}}\chi\\
\end{array}
\right],
\label{eq:learning_direction}
\end{equation}
where $\nabla_{\mathbf{m}^{(n)}} \chi$ is the gradient of the cost function with respect to model parameter $\mathbf{m}^{(n)}$.

One of the most problematic challenges of multiparameter FWI is crosstalk between parameters. 
Crosstalk is due to the fact that different parameters can have a coupled effect on the seismic wavefield and different combination of changes in these parameters can cause similar seismic response \cite[]{OpertoEtAl2013, YangEtAl2018}.
To address this problem, different strategies have been deployed \citep{ metivierEtAl2013, OpertoEtAl2013, lavoue2014two,Fabien_OuelletEtAl2017, KeatingAndInnanen2019}. 
\cite{OpertoEtAl2013} listed four main strategies: the choice of appropriate parameterization, the use of the Hessian as it can measure the level of coupling between parameters, a data driven strategy i.e., using multi-components data, and a sequential inversion where the dominant parameters are estimated before the secondary parameters. 
To further reduce the crosstalk problem,  it is also proposed to scale the gradient \cite[]{KameiEtPratt2013, lavoue2014two}.
Hence equation \ref{eq:learning_direction} can be rewritten as
\begin{equation}
\pmb{\Delta} \mathbf{m} = -\left[ \begin{array}{c}
\nabla_{\mathbf{m}^{(1)}}\chi\\
\xi \nabla_{\mathbf{m}^{(2)}}\chi\\
\beta \nabla_{\mathbf{m}^{(3)}}\chi\\
\end{array}
\right],
\label{eq:learning_direction_scaled}
\end{equation}
where $\xi$ and $\beta$ scale the gradient of different parameter classes.

By using a quasi-Newton method, $\ell$-BFGS, we consider an approximation of the Hessian into the solution \citep{NocedalEtWright2006} which helps reducing crosstalk. 
However the most important point of this study is the chosen parameterization. 
Considering the fact that the time-lapse changes in $P$-wave velocity due to fluid substitution (more affected than $S$-wave velocity and density) are approximately $10\%$ of the baseline model \cite[]{ZhouEtLumley2021},  the challenges of FWI are more pronounced in time-lapse studies than in conventional FWI.
\cite{MaEtAl2016} studied the possibility of deploying TL-FWI for multiparameter acoustic studies.
Although they estimate the model parameters very well, crosstalk prevents from obtaining an acceptable time-lapse image of the subsurface for any of two time-variable parameters.
Thereby, an efficient parameterization is required to minimize the crosstalk in multiparameter TL-FWI. 
Inspired by \cite{HuEtAl2021}, we study the feasibility of performing TL-FWI using the PCS parameterization.
Recalling that the porosity and clay content can be reasonably considered constant with time,  water saturation is the only time-variable parameter.

In the next section, we provide the required equations to adapt the FWI problem to estimate the PCS parameters using Gassmann's equation. 
In this case, the vector of model parameter can be written as 
\begin{equation}
\mathbf{m} = [\phi, C, S_w]^{T}.
\label{eq:model_parameters_pcs}
\end{equation}
Using the aforementioned assumption that porosity and clay content remain constant over time for reservoirs, the changes in water saturation can be the only parameter to be sought in the time-lapse inversion. 

It is worth recalling that the chain rule is used to change the parameters for inversion. 
As an example, $\mathbf{q}$ parameterization, $\mathbf{q}=[q_1, q_2, q_3]^T$, can be changed to a new parameterization, $\mathbf{p}=[p_1, p_2, p_3]^T$,  as 
\begin{equation}
\frac{\partial \chi}{\partial q_i} =\frac{\partial \chi}{\partial p_1}\frac{\partial p_1}{\partial q_i} + \frac{\partial \chi}{\partial p_2}\frac{\partial p_2}{\partial q_i} + \frac{\partial \chi}{\partial p_3}\frac{\partial p_3}{\partial q_i},
\label{eq:general_chain_rule}
\end{equation}
for $i \in (1, 2, 3)$.
FWI can be performed in any parameterization while the forward modeling is implemented using equation \ref{eq:forward_modeling}.  

This study is carried out using PyFWI \cite[]{mardan2023pyfwi} which is an open-source package that implements FWI in terms of DV parameterization.
In the next section, we present the methodology to transfer the earth model from PCS parameterization to DV parameterization for forward modeling.
In Appendix \ref{app:grad_dv2pcs}, we present the required equations to obtain the gradient of the cost function in terms of PCS parameterization for performing the inversion.

\subsection*{Porous media homogenization}
To perform TL-FWI with a rock-physics parameterization, a rock-physics model is needed to map the rock-physics properties to the elastic properties. 
Although there are numerous empirical rock-physics models that satisfy this requirement, we employ Gassmann's equation \citep{gassmann1951} which is valid for most of consolidated rocks \citep{DupuyEtAl2016_1}.
Gassmann's relation is used to compute the variation of bulk modulus during the fluid substitution.
In practice, Gassmann's model requires some information on the fluid and mineral properties.
In the context of TL-FWI, this information should be available as time-lapse FWI is usually performed in well studied fields.

The effective properties of the fluid (bulk modulus, $K$, and density) can be calculated using weighted arithmetic averages \citep{voigt1889ueber},
\begin{equation}
\begin{aligned}
K_f &= S_wK_w + (1 - S_w)K_h,\\
\rho_f &= S_w \rho_w + (1 - S_w)\rho_h,
\end{aligned}
\label{eq:effective_fluid}
\end{equation}
where the subscripts $f$, $w$, and $h$ respectively denote the effective property of the fluid, water, and hydrocarbon.

The density of the solid frame of rock can be calculated using weighted average of the density of the grains as
\begin{equation}
\begin{aligned}
\rho_s = & C \rho_{c} + (1 - C) \rho_q,
\end{aligned}
\end{equation}
where subscripts $s$, $c$, and $q$ denote solid medium, clay, and quartz, respectively. 
The bulk and shear moduli of the solid can be estimated using Voigt-Reuss-Hill method  
\begin{equation}
\begin{aligned}
K_{s}=&\frac{1}{2}\left[\left(C K_{c}+(1-C) K_{q}\right)\right.\\
+&\left.\left(\frac{1}{C / K_{c}+(1-C) / K_{q}}\right)\right], \\
G_{s}=& \frac{1}{2}\left[\left(C G_{c}+(1-C) G_{q}\right)\right.\\
&\left.+\left(\frac{1}{C / G_{c}+(1-C) /G_{q}}\right)\right].
\end{aligned}
\end{equation}
The effective dry moduli of the rock, $K_D$ and $G_D$ can be computed using  various effective medium theories \cite[]{berryman1995mixture,mavko2020rock}. 
Following \cite{pride2005}, we estimate the dry moduli with
\begin{equation}
\begin{aligned}
K_{D} &=K_{s} \frac{1-\phi}{1+c s \phi}, \\
G_{D} &=G_{s} \frac{1-\phi}{1+\frac{3}{2} c s \phi},
\end{aligned}
\label{eq:drained}
\end{equation}
where $cs$ is the general consolidation parameter which defines the degree of consolidation among grains.  
This parameter can be considered as a free parameter in predicting velocities \cite[]{lee2005proposed}. 
 For further studies about $cs$, readers are referred to \cite{lee2005proposed} where the author discusses the importance of this parameter and how it can be calculated.

Finally, the effective properties of the porous medium is calculated  with the following expressions \cite[]{gassmann1951, voigt1889ueber},
 \begin{equation}
 \begin{aligned}
&K =\frac{\phi K_{D}+\left(1-(1+\phi) K_{D} / K_{s}\right) K_{f}}{\phi(1+\Delta)},\\
&G = G_D,\\
&\rho  = (1 - \phi) \rho_s + \phi \rho_f,
\end{aligned}
\label{eq:undrained}
\end{equation}
where
\begin{equation}
\Delta=\frac{1-\phi}{\phi} \frac{K_{f}}{K_{s}}\left(1-\frac{K_{D}}{(1-\phi) K_{s}}\right) = \frac{1-\phi}{\phi} \frac{K_{f}}{K_{s}}\left(1-\frac{1}{1+cs \phi}\right).
\label{eq:delta}
\end{equation}
The elastic properties can be calculated using equation \ref{eq:undrained} as
\begin{equation}
\begin{aligned}
&V_P = \sqrt{\frac{K+ \frac{4}{3}G}{\rho}},\\
&V_S = \sqrt{\frac{G}{\rho}},\\
&\rho = \rho.\\
\end{aligned}
\label{eq:elastic_properties}
\end{equation}
For a detailed discussion on Gassmann's model and the provenance of equations \ref{eq:effective_fluid}-\ref{eq:elastic_properties} see \cite{DupuyEtAl2016_1}.


\subsection*{Time-lapse full-waveform inversion}
\begin{figure}[ht]
\begin{center}
\includegraphics[width=1\textwidth]{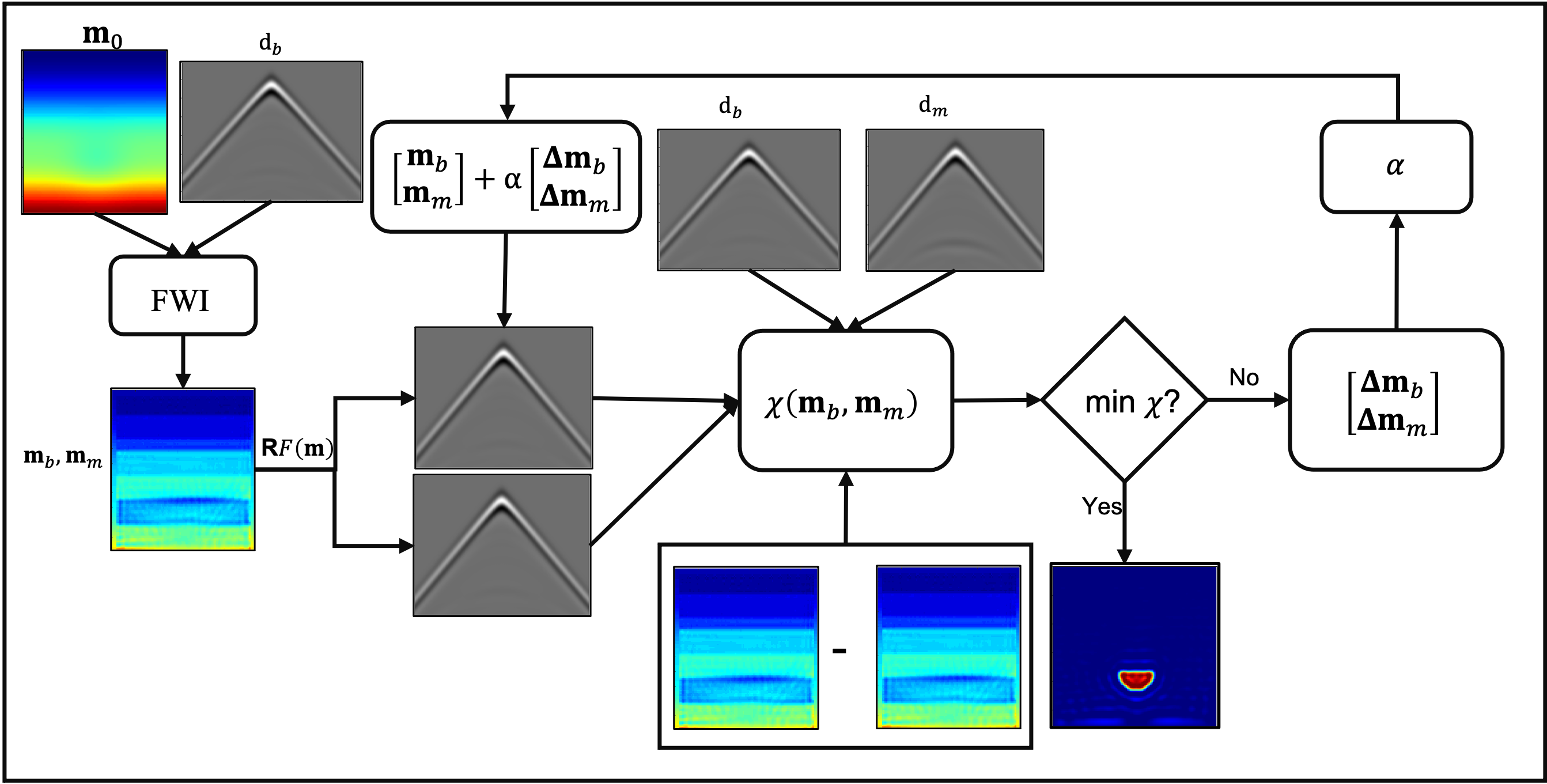}
\caption{Flowchart of simultaneous TL-FWI.  After performing FWI on baseline data,  a joint time-lapse FWI inverts baseline and monitor data where the inversion process is regularized by the difference between estimated baseline and monitor models.
{\label{fig:flowchart}}%
}
\end{center}
\end{figure}

Time-lapse FWI can be performed by considering different combinations of FWI runs.
These combinations can be performed sequentially such as for the cases of the cascaded, cross-updating, and weighted-average methods \cite[]{WatanabeEtAl2005, MaharramovEtAl2016, MardanEtAl2022wa_eage, MardanEtAl2022wa_geophysics}, or in parallel as in the case of the independent and central-difference methods \cite[]{PlessixEtAl2010, ZhouEtLumley2021}.
In addition to these methods, \cite{MaharramovEtAl2016} proposed a joint inversion of seismic data from different vintages by introducing the cost function as
\begin{equation}
\begin{split}
\chi_{tl}(\mathbf{m}_b, \mathbf{m}_m) =& \|\mathbf{Ru}(\mathbf{m}_b)- \mathbf{d}_{b}\|_2^2 + \|\mathbf{Ru}(\mathbf{m}_m)- \mathbf{d}_m\|_2^2\\ 
&+ \delta  \|\mathbf{m}_m - \mathbf{m}_b\|_2^2,
\end{split}
\label{eq:cost_function_tl}
\end{equation}
where subscripts $b$ and $m$ denote baseline and monitor models.
The difference between the estimated baseline and monitor models is used as a regularization term and the weight of this term in the cost function is controlled by parameter $\delta$ which is constant during the inversion.
This parameter, $\delta$, can vary for different surveys and should be picked by trial and error. 
More details on this parameter can be found in \cite{MaharramovEtAl2016}.
The flowchart of this method (hereafter called simultaneous TL-FWI) is presented in Figure \ref{fig:flowchart}.
Simultaneous TL-FWI gets an initial model and observed seismic data from two vintages. 
In the first step, the baseline data are inverted and the estimated baseline is used as the initial model for time-lapse FWI.
TL-FWI minimizes the residuals between the estimated and observed data from the baseline and monitor models regularized by the difference between the estimated models. 

Simultaneous TL-FWI is one of the most accurate strategies for time-lapse studies \citep{MardanEtAl2022wa_geophysics}.  
Incorporating a regularization term for minimizing the difference between two estimates (baseline and monitor models) helps this method to produce less artifacts.

The methodology we propose is as follows. 
We first estimate the baseline model, using standard FWI in the PCS parametrization. 
Then we perform a single-parameter inversion of $S_w$ with jointly inverting the baseline and monitor data (equation \ref{eq:cost_function_tl}). 
We justify the last step by making the hypothesis that porosity and clay content remains constant after fluid injection.
In the case of inversion with the DV parameterization, the TL-FWI is a multi-parameter problem. Furthermore, it requires the additional step of inverting the elastic properties to obtain $S_w$.
In this work, we use $\ell$-BFGS to do so, which minimizes the following cost function
\begin{equation}
\chi_{rp} = \frac{1}{2}\|F(\phi, C, S_w) - \mathbf{m}_{DV}\|^2,
\label{eq:cost_function_rp}
\end{equation}
where $F$ is a function that calculates $V_P$, $V_S$, and $\rho$ based on equations \ref{eq:effective_fluid}-\ref{eq:elastic_properties} and $ \mathbf{m}_{DV}$ is vector of estimated elastic properties using the DV parameterization.

\section*{Numerical analysis}
In this section we assess the feasibility of the proposed strategy with two synthetic models.
For both analysis, FWI is performed to estimate all model properties (elastic or rock-physics properties) simultaneously. 
In this regard, the multi-scale strategy \cite[]{BunksEtAl1995} is employed to avoid cycle-skipping.
 We filter the observed and estimated data using a lowpass Butterworth filter with cut-off frequencies of 10, 20, 25, 35, 40, and 55 Hz.
The source for forward modeling is a Ricker wavelet with a central frequency of 30 Hz.
Perfectly matched layers (PML) are used at the boundaries of the models to minimize waves reflected from the boundaries \cite[]{Berenger1994}. 
We use the Tikhonov regularization (equation \ref{eq:tikhonov}) in the $x$-direction ($\mathbf{B}_z = 0$) to decrease the ill-posedness of FWI problem and make the estimated models smoother in the $x$-direction.

For numerical studies, we use two synthetic models.
For the first experiment, a simple layered model is used.
Then, TL-FWI is performed using the Marmousi model which is a more realistic case.
During the inversion, the gradient is scaled using equation \ref{eq:learning_direction_scaled}.  
The scaling weights $\xi$ and $\beta$ are obtained by trial and error, by performing FWI runs on the simple layered model.
Then we use the same weights for the Marmousi model.
The rock-physical properties used in this study are shown in Table \ref{tbl:rock_properties}.

\begin{table}[h!]
\centering
\begin{tabular}{lccc} 
 \hline
 {} & Bulk modulus (GPa) & Shear modulus (GPa) & Density (g/cm$^3$) \\ [0.5ex] 
 \hline\hline
Quartz      &         37.00 &          44.00 &     2.65 \\
Clay        &         21.00 &          10.0 &     2.55 \\
Water       &          2.25 &           0.00 &     1.00 \\
Gas &          0.04 &           0.00 &     0.10 \\
 \hline
\end{tabular}
\caption{Elastic properties in this study for minerals and fluids where $cs$  is set to 20 \cite[]{DupuyEtAl2016_2, HuEtAl2021}.}
\label{tbl:rock_properties}
\end{table}

\subsection*{Layered model}
\begin{figure}[ht]
\begin{center}
\includegraphics[width=1\textwidth]{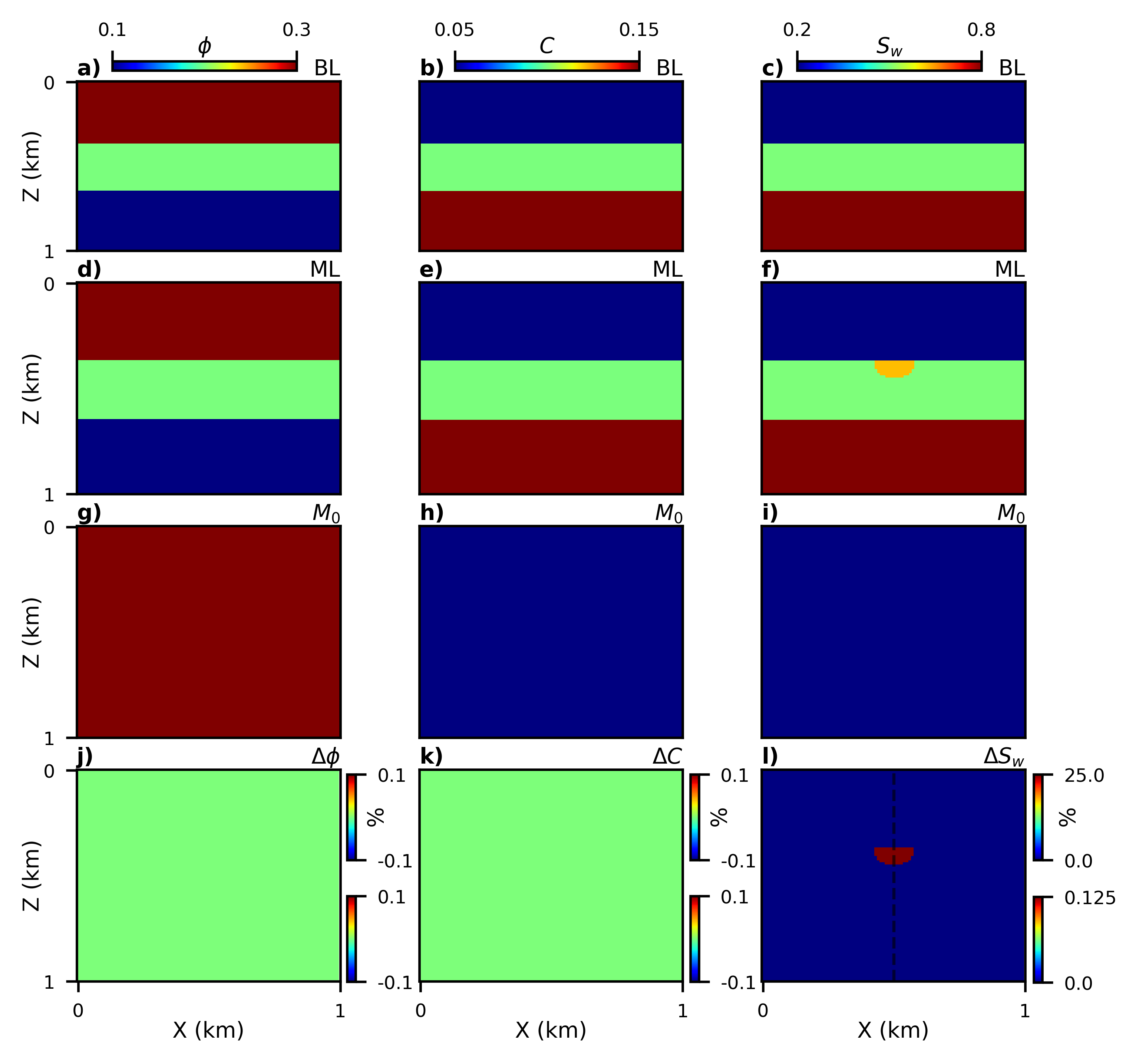}
\caption{Rock-physics properties of the layered model. (a-c) True baseline, (d-f) true monitor, and (g-i) initial models. (j-l) Time-lapse changes between porosity (a and d), clay content (b and e), and water saturation (c and f). The time-lapse model is presented with two color scales which show the changes in value and percentage. Dashed line in (l) is used for 1D profile in Figure~\ref{fig:model2_fwi_bl_1d} and \ref{fig:model2_tlfwi_1d}.
{\label{fig:model2}}%
}
\end{center}
\end{figure}

\begin{figure}[ht]
\begin{center}
\includegraphics[width=1\textwidth]{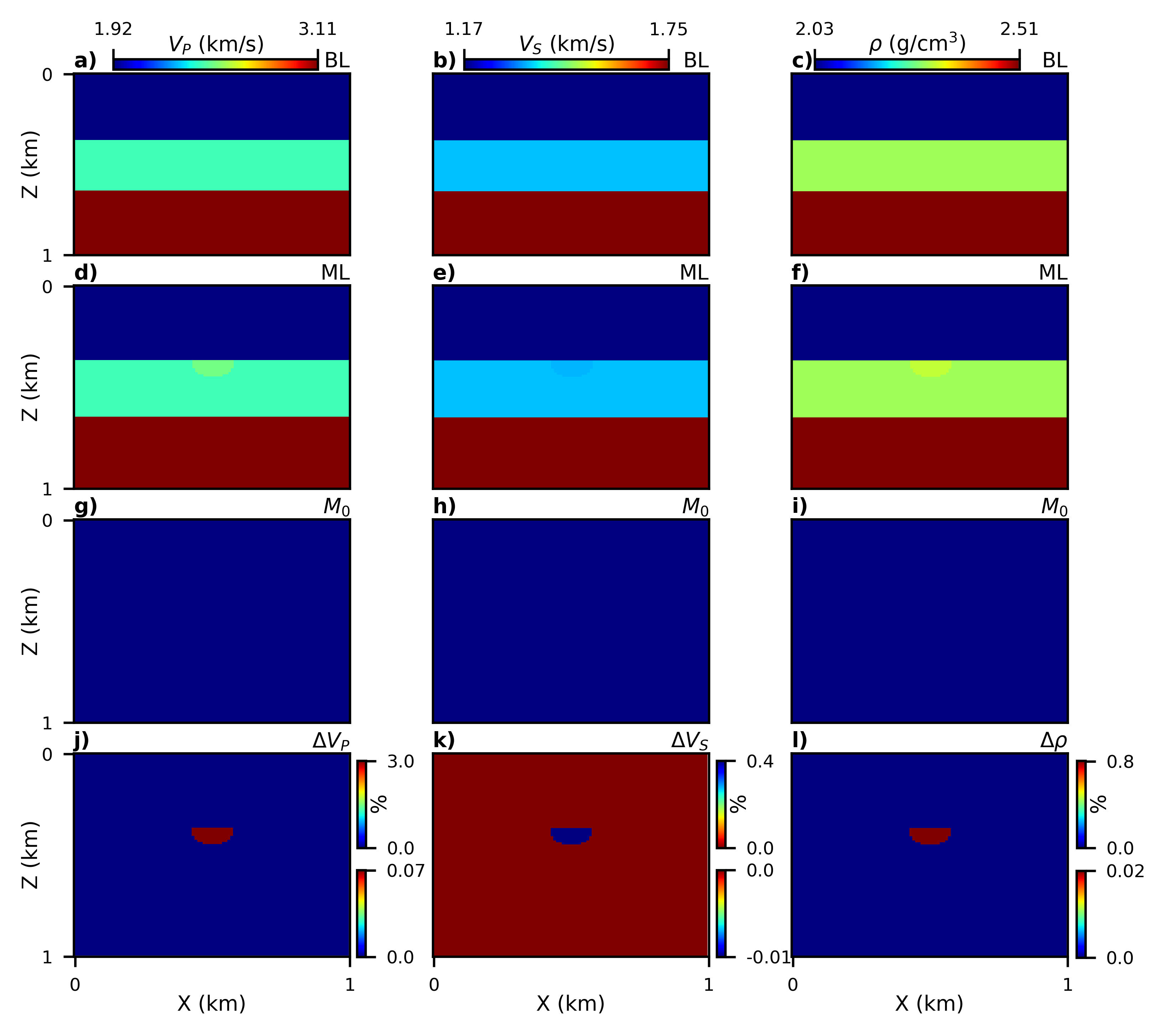}
\caption{Elastic properties of the layered model. (a-c) True baseline, (d-f) true monitor, and (g-i) initial models. (j-l) Time lapse changes between $V_P$ (a and d), $V_S$ (b and e), and $\rho$ (c and f). The time-lapse model is presented with two color scales which show the changes in value and percentage.
{\label{fig:model2_dv}}%
}
\end{center}
\end{figure}

To investigate the performance of the methodology, a simple model with three flat reflectors is first used (Figure~\ref{fig:model2}).
Figure~\ref{fig:model2}a-\ref{fig:model2}c shows the true models for porosity, clay content and water saturation for the baseline and Figure~\ref{fig:model2}d-\ref{fig:model2}f shows the true monitor model.
The initial model of the inversion is shown in Figure~\ref{fig:model2}g-\ref{fig:model2}i.
Finally, the true time-lapse model is shown in Figure~\ref{fig:model2}j-\ref{fig:model2}l.
The elastic properties of this model are shown in Figure~\ref{fig:model2_dv}.
Comparing Figure~\ref{fig:model2}j-\ref{fig:model2}l with Figure~\ref{fig:model2_dv}j-\ref{fig:model2_dv}l shows that with a $25\%$ increment of water saturation, P-wave velocity and density increase by $3\%$ and $0.8\%$, respectively, while the S-wave velocity decreases by $0.4\%$.


\begin{figure}[ht]
\begin{center}
\includegraphics[width=1\textwidth]{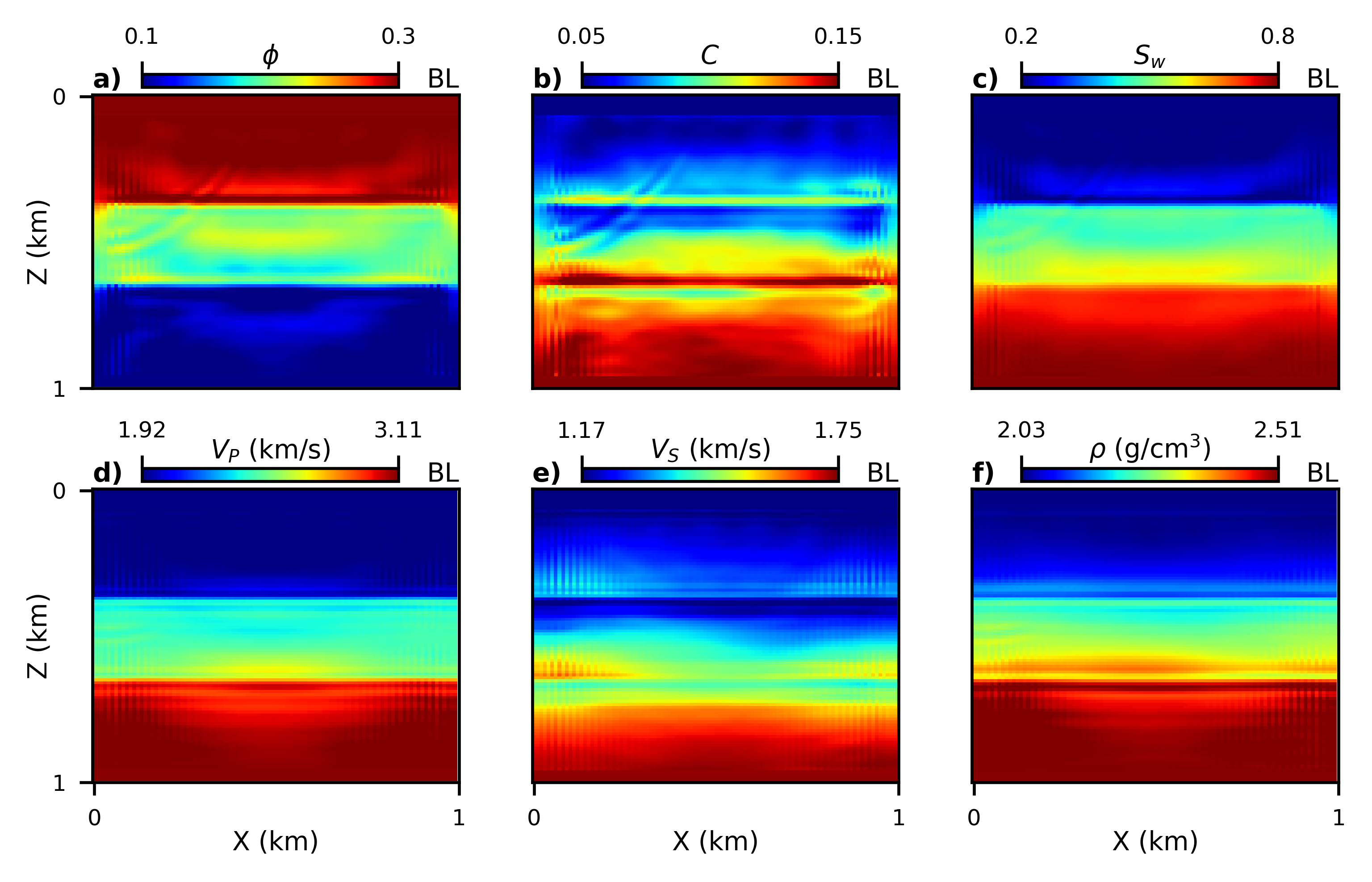}
\caption{The estimated baseline model of (a) porosity, (b) clay content, (c) water saturation, (d) $P$-wave velocity, (e) $S$-wave velocity, and (f) density estimated using the PCS parameterization.
{\label{fig:model2_fwi_bl}}
}
\end{center}
\end{figure}

\begin{figure}[ht]
\begin{center}
\includegraphics[width=1\textwidth]{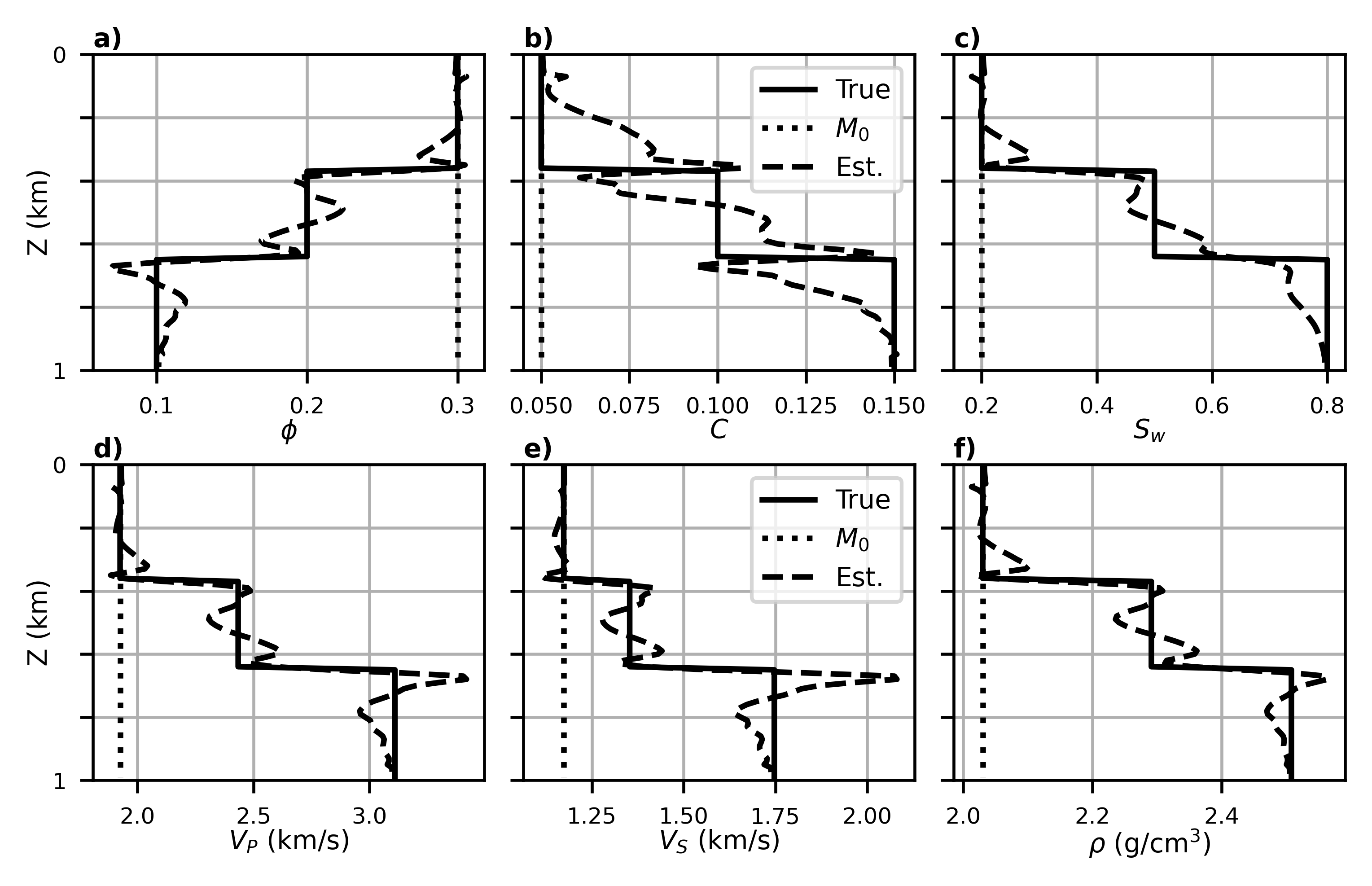}
\caption{1D assessment of the results for (a) porosity, (b) clay content, (c) water saturation, (d) $P$-wave velocity, (e) $S$-wave velocity,  and (f) density at $X=0.5$ km (center of the 2D section). True value is shown with solid line, estimated model is shown with dashed line and dotted line presents the initial model.
{\label{fig:model2_fwi_bl_1d}}
}
\end{center}
\end{figure}

To perform this study, seven isotropic sources are used on the surface for forward modeling.
Receivers are located on the surface and in two imaginary wells at both sides of the model.
Noise-free pressure data are inverted to recover the time-lapse anomaly.
The rock-physics properties estimated by using Gassmann's equation are presented in Figure~\ref{fig:model2_fwi_bl}a-\ref{fig:model2_fwi_bl}c.    
The parameters in this model are used to obtain an indirect estimate of $P$- and $S$-wave velocities as well as density, as shown in Figure~\ref{fig:model2_fwi_bl}d-\ref{fig:model2_fwi_bl}f. 
A 1D plot obtained at $X = 0.5$ km (dashed line in Figure~\ref{fig:model2}l) is also shown in Figure~\ref{fig:model2_fwi_bl_1d} for both rock-physics properties, Figure~\ref{fig:model2_fwi_bl_1d}a-\ref{fig:model2_fwi_bl_1d}c, and elastic properties, Figure~\ref{fig:model2_fwi_bl_1d}d-\ref{fig:model2_fwi_bl_1d}f.

Results of the first step obtained with this simple synthetic model shows that FWI and the proposed formulation have the potential to recover the rock-physics properties.  
However, reconstructing the clay content seems problematic and the interfaces in this parameter are not recovered sharply (Figure~\ref{fig:model2_fwi_bl}b and \ref{fig:model2_fwi_bl_1d}b).  
Although \cite{HuAndInnanen2021} showed that the $C$ model can be estimated better by considering a parameter-relation regularization,  this type of regularization is not taken into account in this study. 
The parameter-relation regularization allows  us to regularize the inversion based on the relation between two parameters in the field.
So in general, we can expect that the parameter-relation and prior-model regularization  \cite[]{AsnaashariEtAl2013} should improve the accuracy of the estimated clay content.
These two regularization methods are provided in Appendix \ref{app:regularization} for interested readers.

\begin{figure}[ht]
\begin{center}
\includegraphics[width=1\textwidth]{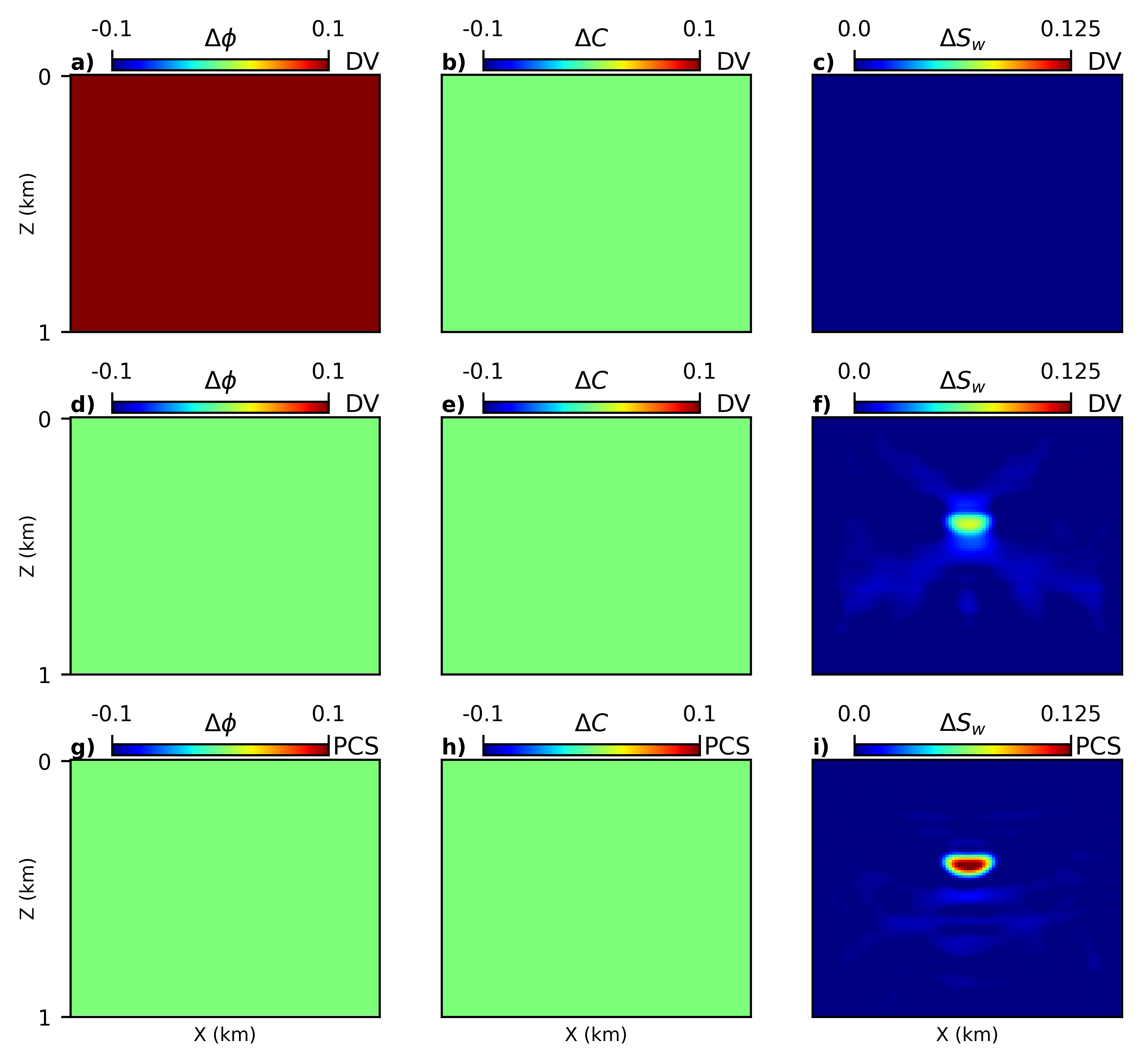}
\caption{Estimated time-lapse model for $\phi$, $C$, and $S_w$ with (a-c) the DV parameterization while all model parameters can vary with time, (d-f) the DV parameterization while $S_w$ is the only time-dependent parameter, and (g-i) the PCS parameterization. 
{\label{fig:model2_tl_rp}}%
}
\end{center}
\end{figure}

After estimating the baseline model (Figure~\ref{fig:model2_fwi_bl}), TL-FWI can be performed using equation \ref{eq:cost_function_tl}.
We first compare the efficiency of the DV and PCS parameterization to monitor the rock-physics properties.
Using the DV parameterization, changes in terms of the rock-physics properties are estimated in two steps.
We first estimate the elastic properties of the baseline and the monitor models using equation \ref{eq:cost_function_tl}.
In the second step, equation \ref{eq:cost_function_rp} is employed to estimate the rock-physics properties of the baseline and the monitor models and differences are presented in Figure~\ref{fig:model2_tl_rp}a-\ref{fig:model2_tl_rp}c.
This is a naive method to perform multiparameter TL-FWI and leads to overestimation of porosity and inaccurate time-lapse estimate.
We cannot get an accurate time-lapse image using this method due to the crosstalk between parameters.
It is worth reminding that the crosstalk problem is more severe in time-lapse inversion, because the magnitude of changes in reservoirs is comparable to generated artifacts.
To improve the results, we follow another strategy.
In this strategy, the rock-physics properties of the baseline are estimated using equation \ref{eq:cost_function_rp}.
Then, by assuming that porosity and clay content are not variable with time,  the elastic properties of the monitor model are inverted to estimate the saturation. 
For this purpose, the estimated rock-physics model of the baseline is used as initial model to estimate the rock-physics properties of the monitor model.
Despite the inversion of the baseline model, we perform single-parameter inversion for estimating the rock-physics properties of the monitor model during which only $S_w$ is updated.
This strategy leads to the time-lapse estimate presented in Figure \ref{fig:model2_tl_rp}d-\ref{fig:model2_tl_rp}f where a better time-lapse image of the saturation is obtained.
This strategy is used hereafter in this study to compute the time-lapse rock-physics properties using the DV parameterization.
By using the PCS parameterization and considering time-independent porosity and clay content, direct TL-FWI is performed to only monitor the water saturation. 
The estimated time-lapse models of porosity, clay content, and water saturation are presented in Figure~\ref{fig:model2_tl_rp}g-\ref{fig:model2_tl_rp}i.
Comparing Figure~\ref{fig:model2_tl_rp} to Figure~\ref{fig:model2}j-\ref{fig:model2}l shows that the PCS parameterization can improve the accuracy of time-lapse studies for monitoring the saturation of the fluids in the subsurface.
The accuracy of the results is measured using root-mean-square error (RMSE) and direct monitoring of $S_w$ using the PCS parameterization leads to a $37.3\%$ higher accuracy in comparison to using the DV parameterization.

\begin{figure}[ht]
\begin{center}
\includegraphics[width=1\textwidth]{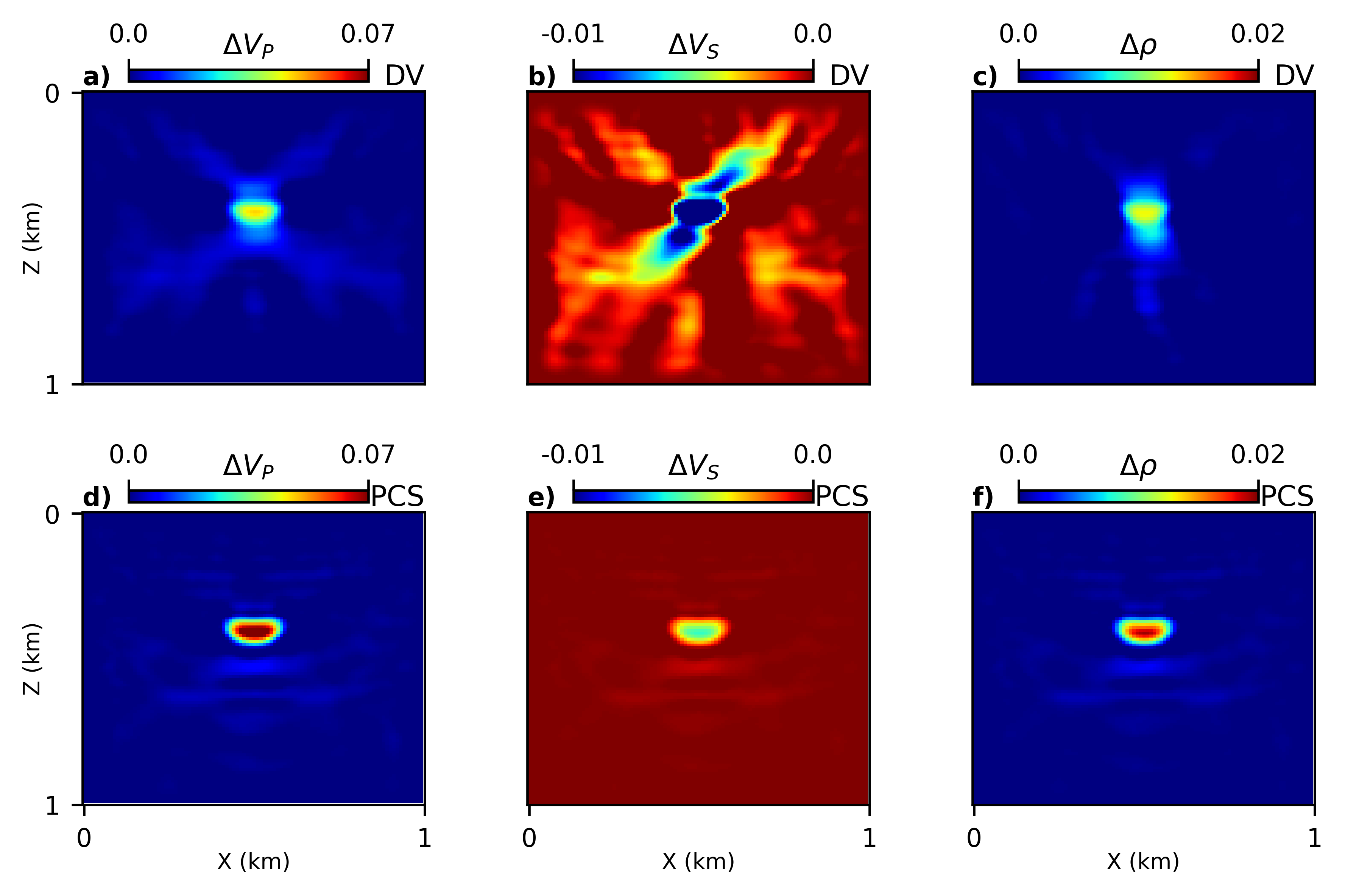}
\caption{Estimated time-lapse model for $V_P$, $V_S$, and $\rho$ with (a-c) the DV parameterization and (d-f) the PCS parameterization. 
{\label{fig:model2_tl_elastic}}%
}
\end{center}
\end{figure}

In addition to the rock-physics properties, the PCS parameterization can be used to recover time-lapse images of the elastic properties.
For this case, we use equation \ref{eq:cost_function_tl} to obtain the baseline and monitor models in terms of porosity, clay content and water saturation.
The results are then used for rock-physics modeling (equation \ref{eq:effective_fluid}-\ref{eq:elastic_properties}) to obtain the elastic properties of the baseline and monitor models.
While the DV parameterization leads to time-lapse images presented in Figure \ref{fig:model2_tl_elastic}a-\ref{fig:model2_tl_elastic}c, the PCS parameterization improves the time-lapse estimates (Figure \ref{fig:model2_tl_elastic}d-\ref{fig:model2_tl_elastic}f).
The DV parameterization does not allow recovering the changes appropriately.
This can be explained by considering that $25\%$ changes in the saturation (Figure \ref{fig:model2}l) is now spread out between three elastic properties and it causes only $3\%$, $0.4\%$ and $0.8\%$ variation respectively in $V_P$, $V_S$, and $\rho$ (Figure \ref{fig:model2_dv}j-\ref{fig:model2_dv}l).
Comparing the estimated time-lapse changes in Figure~\ref{fig:model2_tl_elastic} shows that using the PCS parameterization, we can recover the elastic properties more accurately than using the DV parameterization. 
The time-lapse elastic properties estimated using the PCS parameterization are $53\%$ more accurate than with the DV parameterization.

\begin{figure}[ht]
\begin{center}
\includegraphics[width=1\textwidth]{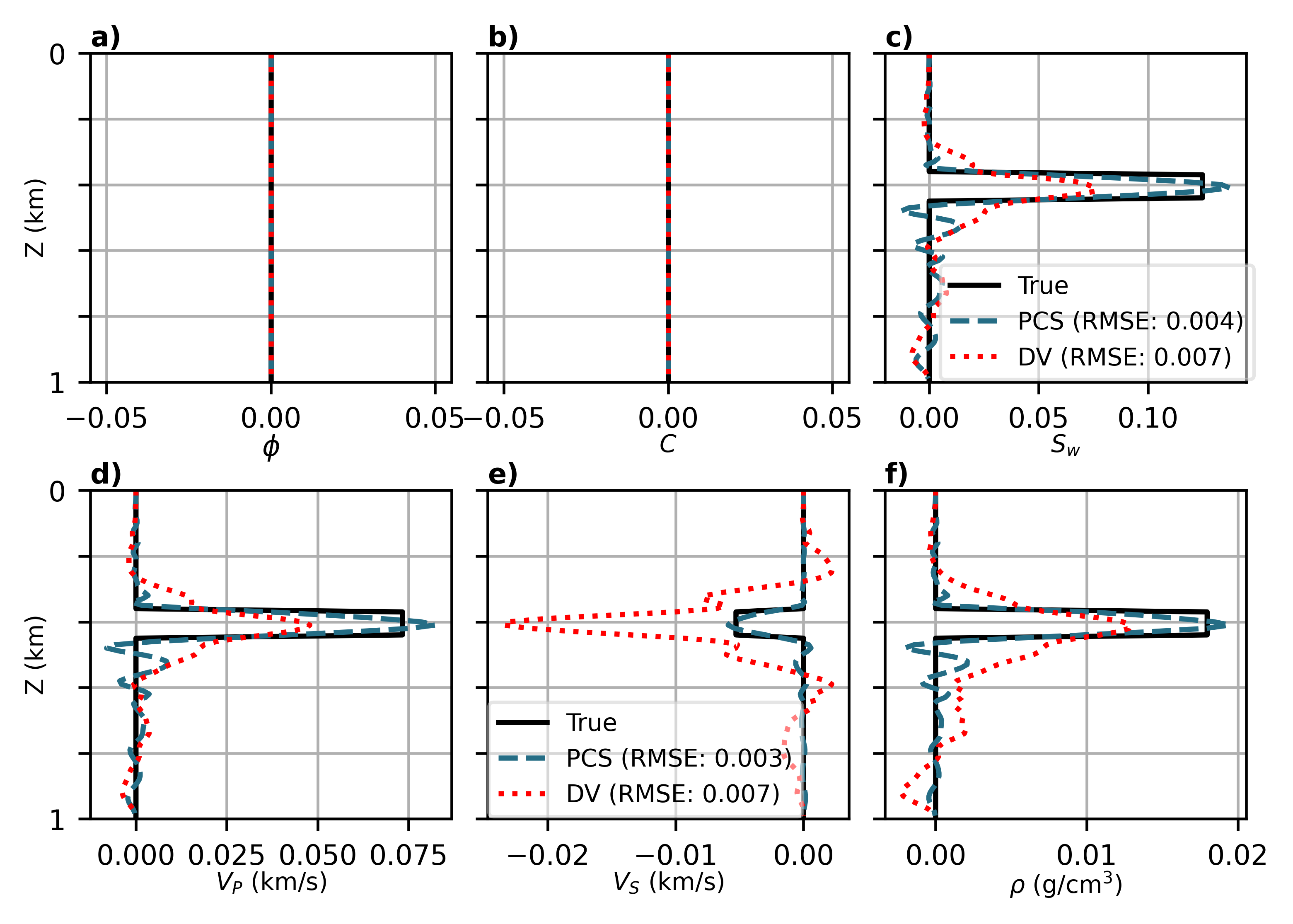}
\caption{1D assessment of the time-lapse estimates for (a) porosity, (b) clay content, (c) water saturation, (d) $P$-wave velocity, (e) $S$-wave velocity,  and (f) density at $X=0.5$ km. True value is shown with solid line, estimated model with PCS parameterization is shown with dashed line and dotted line presents the estimated model with the DV parameterization.
{\label{fig:model2_tlfwi_1d}}%
}
\end{center}
\end{figure}

Figure~\ref{fig:model2_tlfwi_1d}a-\ref{fig:model2_tlfwi_1d}c shows 1D profiles of the true and estimated changes in the rock-physics properties along the dashed line in Figure~\ref{fig:model2}l.
The estimated time-lapse elastic properties from PCS and DV parameterizations are also compared in Figure~\ref{fig:model2_tlfwi_1d}d-\ref{fig:model2_tlfwi_1d}f.
This comparison shows that the results obtained using the DV parameterization (dotted lines) are far from accurate.
For 1D analysis, the PCS parameterization leads to $43\%$ higher accuracy (Figure \ref{fig:model2_tlfwi_1d}a-\ref{fig:model2_tlfwi_1d}c).
For monitoring the elastic properties, the PCS parameterization yields $57\%$ higher accuracy (Figure \ref{fig:model2_tlfwi_1d}d-\ref{fig:model2_tlfwi_1d}f).

\clearpage

\subsection*{Marmousi model}
\begin{figure}[ht]
\begin{center}
\includegraphics[width=1\textwidth]{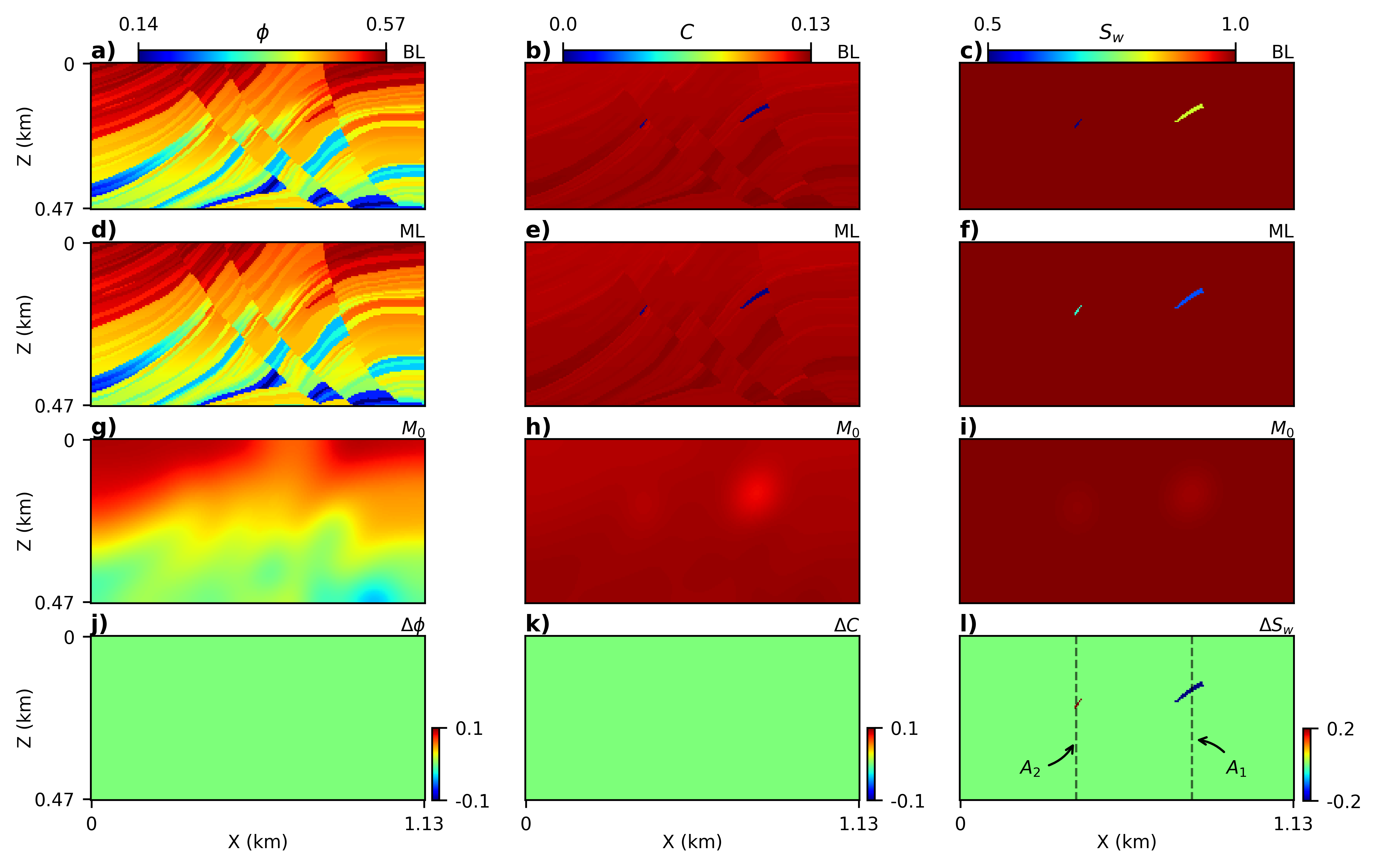}
\caption{Rock-physics properties of the Marmousi model. (a-c) True baseline, (d-f) true monitor, and (g-i) initial models. (j-l) Time-lapse changes between porosity (a and d), clay content (b and e), and water saturation (c and f). Lines A$_1$ and A$_2$ are used for 1D studies in Figure~\ref{fig:model4_tlfwi_1d}.
{\label{fig:model4}}%
}
\end{center}
\end{figure}

In this section, we aim at assessing the methodology using a more realistic model.
For this purpose, a portion of the Marmousi model is chosen (Figure~\ref{fig:model4}).
The true baseline and monitor models are shown in Figure~\ref{fig:model4}a-\ref{fig:model4}f. 
This model contains two sandstone reservoirs.
The monitor model (Figure~\ref{fig:model4}d-\ref{fig:model4}f) is created by reducing the water saturation in the reservoir indicated by line A$_1$ and increasing the water saturation in the reservoir indicated by line A$_2$ (Figure~\ref{fig:model4}l). 

\begin{figure}[ht]
\begin{center}
\includegraphics[width=1\textwidth]{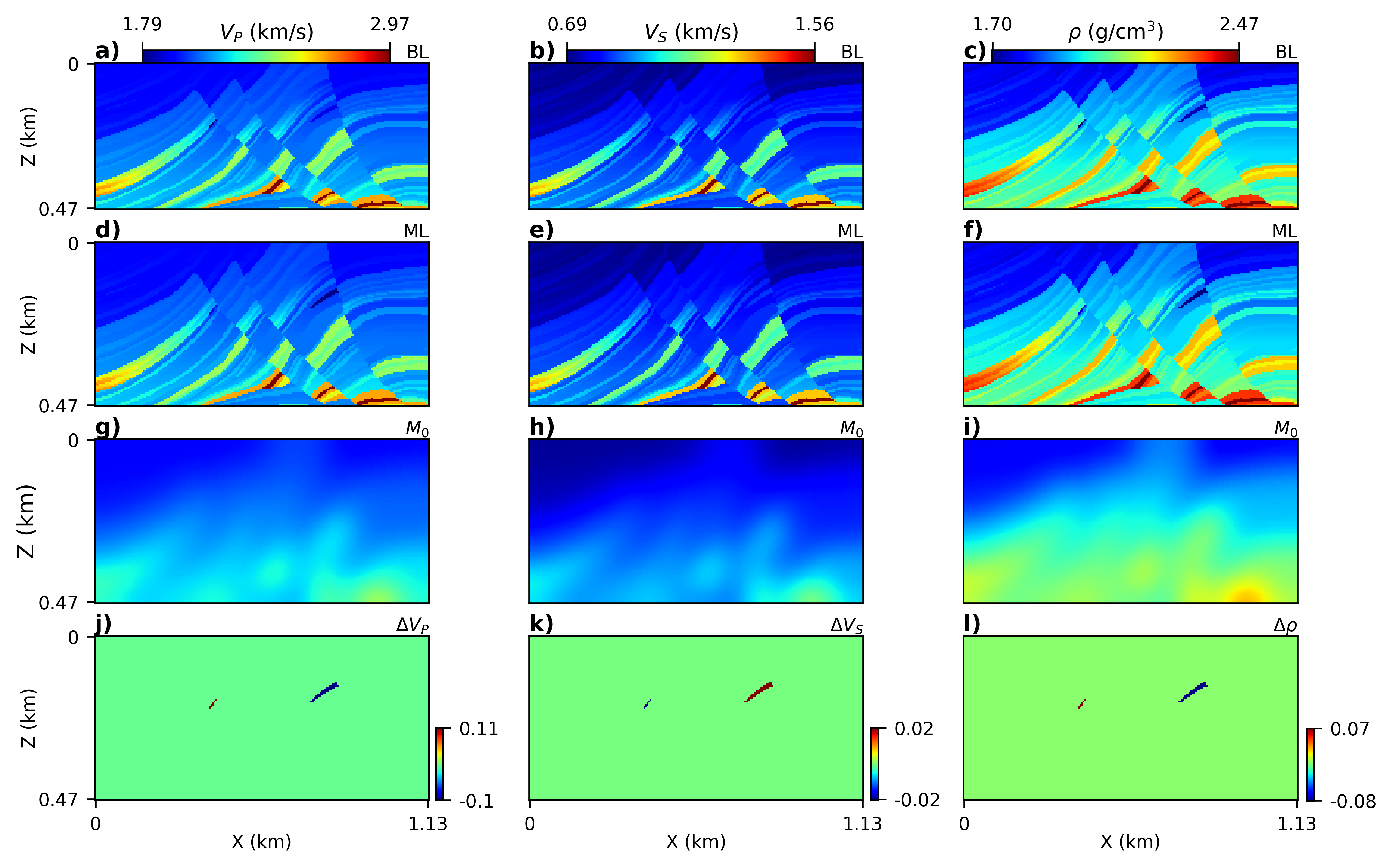}
\caption{Elastic properties of the Marmousi model. (a-c) True baseline, (d-f) true monitor, and (g-i) initial models. (j-l) Time-lapse changes between $V_P$ (a and d), $V_S$ (b and e), and $\rho$ (c and f). 
{\label{fig:model4_dv}}%
}
\end{center}
\end{figure}

\begin{figure}[ht]
\begin{center}
\includegraphics[width=1\textwidth]{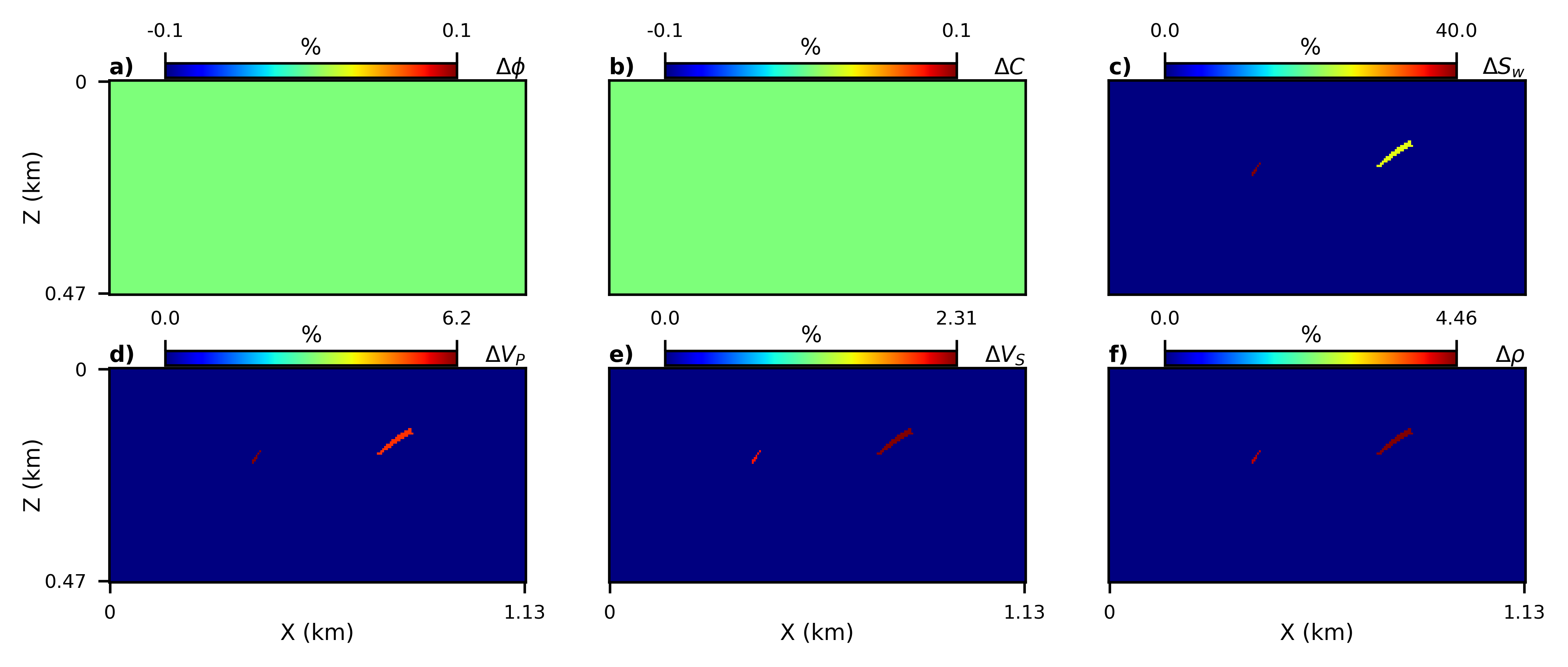}
\caption{Absolute value of percent changes relative to the baseline in terms of studied (a-c) rock-physics and (d-f) elastic properties.
{\label{fig:model4_tl_percentage}}%
}
\end{center}
\end{figure}

The elastic properties of this model are shown in Figure~\ref{fig:model4_dv}.
Basically, the density of the saturated rock increases by replacing the hydrocarbon by water. 
Although the $P$-wave velocity has a reverse relationship with density, the hydrocarbon-water replacement increases the $P$-wave velocity due to the increase of the saturated bulk modulus.
However, the fluid shear modulus is nil and $S$-wave velocity is decreased due to the increase in density.
Figure~\ref{fig:model4}j-\ref{fig:model4}l and Figure~\ref{fig:model4_dv}j-\ref{fig:model4_dv}l show the time-lapse changes in rock-physics and elastic properties, respectively. 
The relative changes are shown in Figure~\ref{fig:model4_tl_percentage}a-\ref{fig:model4_tl_percentage}c and Figure~\ref{fig:model4_tl_percentage}d-\ref{fig:model4_tl_percentage}f for the rock-physics and elastic properties.
In this case, the maximum changes in water saturation occurs in the reservoir A$_2$, which shows a $40\%$ change with respect to the baseline.
This change leads to maximum change of $6.2\%$, $2.31\%$, and $4.46\%$ in elastic properties,  $V_P$, $V_S$, and $\rho$ respectively. 
Hence, we can expect that changes can be better tracked using the PCS parameterization.

\begin{figure}[ht]
\begin{center}
\includegraphics[width=1\textwidth]{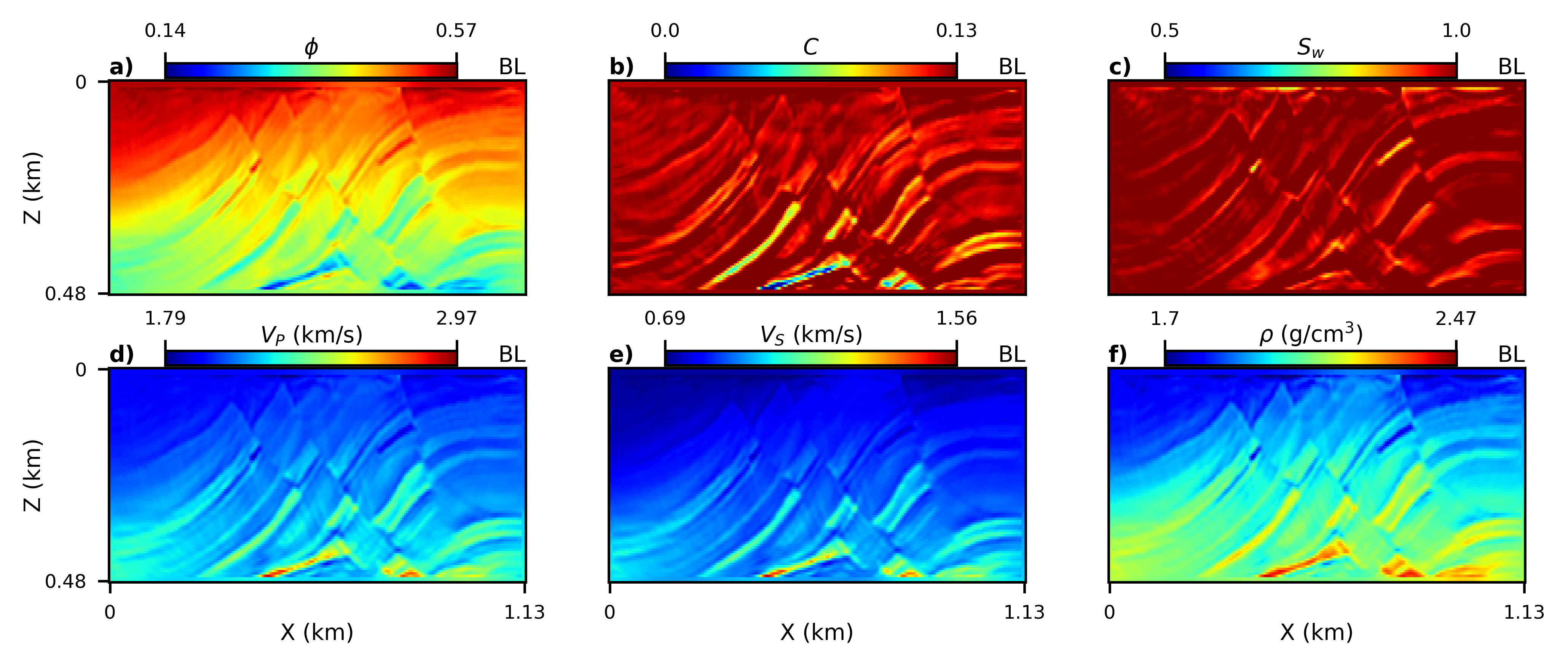}
\caption{The estimated models of (a) porosity, (b) clay content, (c) water saturation, (d) $P$-wave velocity, (e) $S$-wave velocity, and (f) density estimated using the PCS parameterization.
{\label{fig:model4_fwi_bl}}%
}
\end{center}
\end{figure}

\begin{figure}[ht]
\begin{center}
\includegraphics[width=1\textwidth]{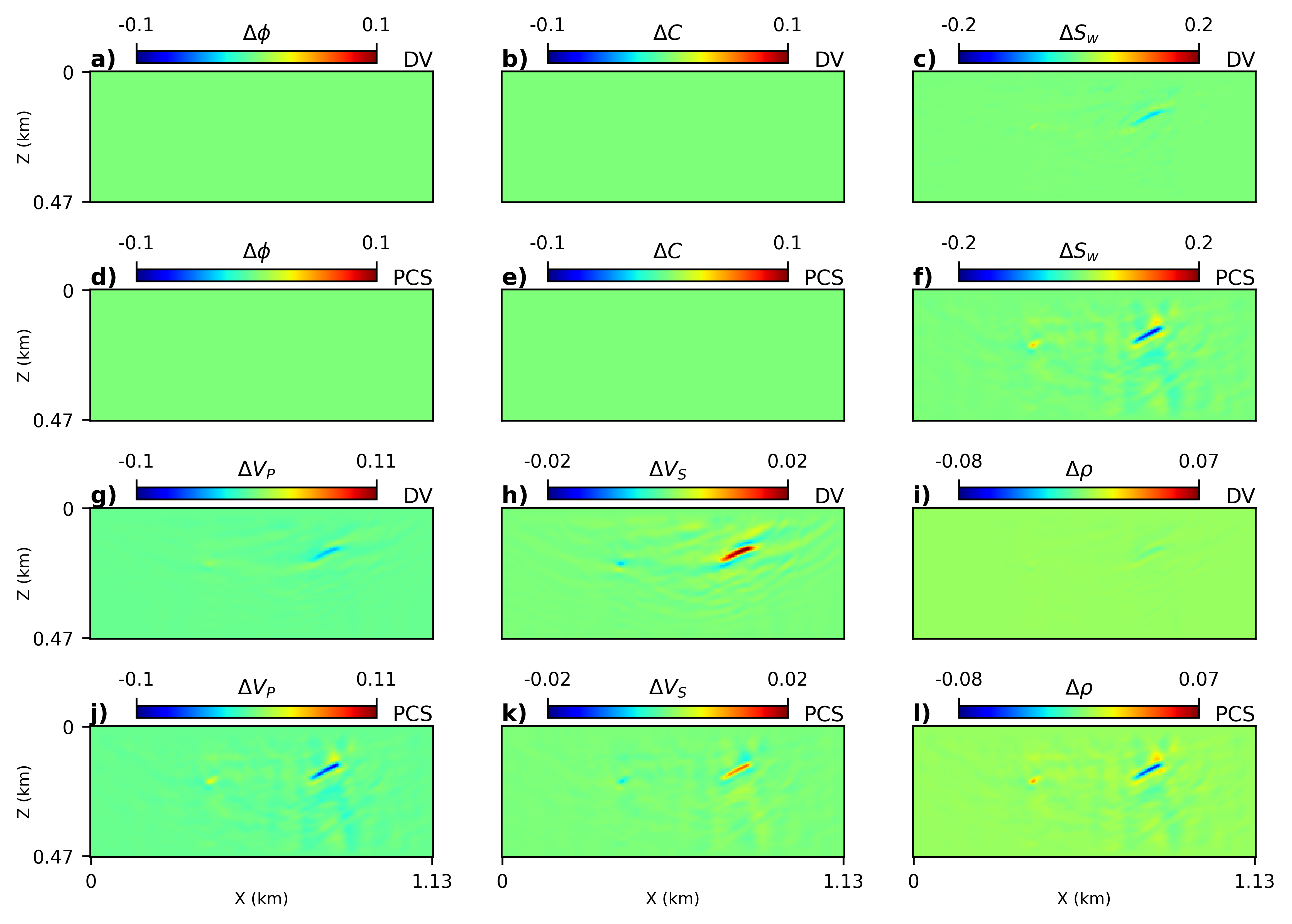}
\caption{Estimated time-lapse model for $\phi$, $C$, and $S_w$ with (a-c) the DV and (d-f) PCS parameterizations. Estimated time-lapse model for $V_P$, $V_S$, and $\rho$ with (g-i) the DV and (j-l) the PCS parameterizations.
{\label{fig:model4_tl}}%
}
\end{center}
\end{figure}

FWI of the baseline data is carried out using the PCS parameterization with values of respectively 5 and 8 for $\xi$ and $\beta$, obtained from the layered model, studied in the previous section.
The initial model for this inversion is presented in Figure~\ref{fig:model4}g-\ref{fig:model4}i.
The results of the inversion of the baseline are presented in Figure~\ref{fig:model4_fwi_bl}.
As the receivers are located only on the surface for this model, it can be seen that the quality of the model update decreases toward the edges. 
Finally, the results of TL-FWI are shown in Figure~\ref{fig:model4_tl}.
As shown in Figure~\ref{fig:model4_tl}a-\ref{fig:model4_tl}c, the DV parameterization allows recovering changes in only reservoir A$_1$.
On the other hand, the time-lapse water saturation anomalies are detected in both reservoirs using the PCS parameterization, as shown in Figure~\ref{fig:model4_tl}d-\ref{fig:model4_tl}f.

Figure~\ref{fig:model4_tl}g-\ref{fig:model4_tl}l shows the changes in elastic properties. 
In this case, the PCS parameterization, shown in Figure~\ref{fig:model4_tl}j-\ref{fig:model4_tl}l, yields significant improvement over the DV parameterization, shown in Figure~\ref{fig:model4_tl}g-\ref{fig:model4_tl}i. 
This improvement leads to $11.7\%$ higher accuracy by using the PCS parameterization.
As can be seen in Figure~\ref{fig:model4_tl}g, the changes in $P$-wave velocity are not recovered in reservoir $A_2$.
Although the time-lapse changes in $S$-wave velocity can be detected, the value in reservoir $A_1$ is overestimated (Figure~\ref{fig:model4_tl}h).
Last but not least, the time-lapse anomalies of density are not detectable from Figure~\ref{fig:model4_tl}i.
Despite inaccuracies in the estimated time-lapse elastic properties, Figure~\ref{fig:model4_tl}j-\ref{fig:model4_tl}l show that the PCS parameterization can lead to a better estimation of elastic properties.

\begin{figure}[ht]
\begin{center}
\includegraphics[width=1\textwidth]{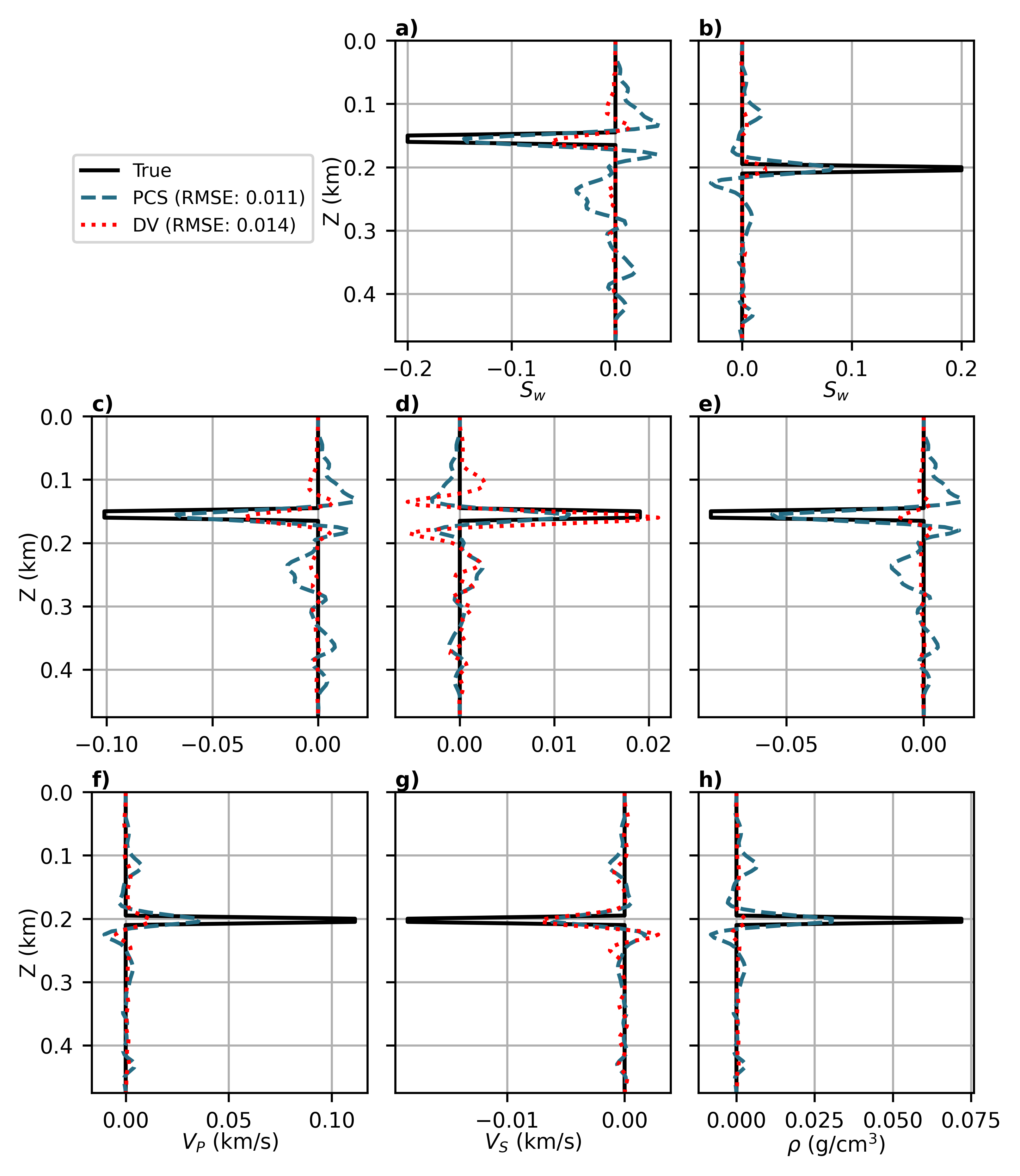}
\caption{1D assessment of the time-lapse estimates for saturation along the line (a) A$_1$, (b) A$_2$. Time-lapse estimates for elastic properties along the line (c-e) A$_1$ and (f-h) A$_2$. True value is shown with solid line, estimated model with the PCS parameterization is shown with dashed line and the dotted line presents the estimated model with the DV parameterization. 
{\label{fig:model4_tlfwi_1d}}%
}
\end{center}
\end{figure}

Finally, a 1D assessment of the results is carried out along lines A$_1$ and A$_2$, shown in Figure~\ref{fig:model4}l.
The 1D presentation of time-lapse changes (Figure~\ref{fig:model4_tlfwi_1d}) shows that TL-FWI with the DV parameterization is not able to recover the changes in the reservoirs appropriately, while using the PCS parameterization can provide accurate time-lapse images of the model for both the elastic and rock-physics properties.
In this 1D analysis, the PCS parameterization leads to $24.51\%$ higher accuracy for estimating the elastic properties and $19.57\%$ for estimating the rock-physics properties.

\section*{Discussion}
The potential of the PCS parameterization for estimating time-lapse changes in elastic and rock-physics properties is shown in this study.
The direct estimation of changes in rock-physics properties provides more accurate results rather than the indirect estimation.
However, the estimates presented in Figure~\ref{fig:model2_tl_rp} show clearly that an effective inversion strategy and optimization method can improve the accuracy of estimated changes using indirect inversion.
As the computation time of the rock-physics modeling is negligible, global inversion methods such as Monte-Carlo methods might be a better choice for the second step of the inversion \citep{DupuyEtAl2016_1}.
However, the direct estimation of rock-physics properties aid to avoid the challenges of picking an appropriate optimization method and parameters for the rock-physics inversion.

The estimated time-lapse changes for the two presented synthetic models show that using the PCS parameterization can also enhance the accuracy for monitoring the elastic properties (Figure \ref{fig:model2_tl_elastic} and Figure~\ref{fig:model4_tl}).
This is due to the fact that using the PCS parameterization, the crosstalk between parameters of different vintages can be avoided.
In this study, we assumed that the only variable parameter with time is fluid saturation.
However, pore pressure is another important parameter that should be considered.
Although the changes in the pore pressure can be negligible in some fields due to the properties such as high permeability \citep{dupuy2021combined}, further studies should be conducted to analyze the efficiency of simultaneous monitoring of saturation and pore pressure.

In this work, we have assumed a simple rock-physics model in which only saturation effects on elastic properties are modeled. 
In reality, this model will contain errors in describing the actual subsurface properties either from the calibration and subsurface assumptions or even the model itself. 
However, time-lapse studies are carried out in well studied fields. 
The calibration of the rock-physics model should not pose too much of a problem in this scenario. 
Besides, the geological information and well-log data of these fields can be employed to regularize the inversion problem and increase the accuracy of FWI results.

\section*{Conclusions}
Time-lapse seismic data have been proven as an important source of information for reservoir monitoring. 
Full-waveform inversion is a powerful tool that relies on seismic data to recover the subsurface properties. 
However this method generally suffers from crosstalk between parameters in multiparameter studies.
This problem becomes more dramatic in time-lapse studies, because the time-lapse changes can be caused by any combination of different parameters in different vintages.

In this study, we formulated time-lapse full waveform inversion in terms of porosity, clay content, and water saturation using Gassmann's equation, to assess its performance for reducing crosstalk and improve estimates of saturation changes.
Using this formulation, we can rely on the assumption that water saturation is the only variable parameter that changes in time and thereby, it is possible to avoid crosstalk caused by time-lapse study. 
We showed that the PCS parameterization improves the time-lapse estimate.
By using Gassmann's equation and the new parameterization, it is also shown that the time-lapse estimate in elastic properties can be obtained indirectly with higher accuracy ($11.7\%$ higher accuracy in case of the Marmousi model).

\section*{Acknowledgment}
We would like to thank François Lavoué for valuable discussions on multiparameter full-waveform inversion.
This project was supported by a NSERC Discovery Grant to BG (RGPIN-2017-06215).
A.Mardan was also partially supported by SEG/Landmark Scholarship.

\section*{Data availability statement}
Data associated with this research are available and can be obtained
by contacting the corresponding author.

\clearpage

\bibliographystyle{seg}  
\bibliography{direct_monitoring_arxiv.bib}  

\newpage
\begin{appendices}
\section{Regularization in seismic FWI}
\label{app:regularization}
By taking advantage of Tikhonov regularization, smoothness of the estimated model can be achieved \cite[]{AsnaashariEtAl2013}.
This regularization can be presented as
\begin{equation}
\chi_1 = \lambda_1\left( \mathbf{m}^T \mathbf{B}_x^T\mathbf{B}_x\mathbf{m} +  \mathbf{m}^T \mathbf{B}_z^T\mathbf{B}_z\mathbf{m} \right),
\label{eq:tikhonov}
\end{equation}
where $\mathbf{B}$ and $\lambda$ denote respectively the first-order spatial derivative operator and regularization hyperparameter.
The subscripts $x$ and $z$ specify the direction.

Total variation regularization (TV) has also shown promising results in FWI to decrease artifacts and preserve the edges and the discontinuities \cite[]{anagaw2012edge, esser2018total}. 
This regularization can be achieved by using 
\begin{equation}
\chi_2 = \lambda_2 \| \sqrt{(\mathbf{B}_x\mathbf{m})^2 +  (\mathbf{B}_z\mathbf{m})^2}\|_1.
\label{eq:tv}
\end{equation}

Also, \cite{AsnaashariEtAl2013} state that where nonseismic information exists, it should be used as prior information for FWI inversion. 
Hence, they proposed regularizing the cost function using the prior information ($ \mathbf{m}_p$) as 
\begin{equation}
\chi_3 =  \lambda_3\|  \mathbf{W}_m(\mathbf{m} -  \mathbf{m}_p)\|_2^2,
\label{eq:prior}
\end{equation}
where $\mathbf{W}_m$ is a weighting operator in the model space. 

Finally, in the case of multiparameter FWI, if information about the relationship between different parameters exists, then regularization can be considered to force the optimization algorithm to follow \cite[]{HuAndInnanen2021},
\begin{equation}
\chi_4 =\lambda_4\| \mathbf{m}_1 - f(\mathbf{m}_2) \|_2^2,
\label{eq:parameter_relation}
\end{equation}
where $f$ is a function to estimates $\mathbf{m}_1$ from  $\mathbf{m}_2$.

\section{The gradient of cost function in PCS parameterization}\label{app:grad_dv2pcs}
In equation \ref{eq:undrained}, we showed how to calculate the elastic properties of the porous medium using $K$, $G$, and $\rho$.
To perform FWI with PCS parameterization, it is required to estimate the gradient of these parameters in terms of PCS parameterization.
Gradient of $K$, $G$, and $\rho$ with respect to $\phi$ can be obtained as
\begin{equation}
\begin{aligned}
\begin{split}
\frac{\partial K}{\partial \phi} &= \frac{K_D + \phi \frac{\partial K_D}{\partial \phi} - \frac{K_f}{K_s}\left(K_D + (1 + \phi)\frac{\partial K_D}{\partial \phi}\right)}{\phi (1 + \Delta)} - \\
&  \frac{\left[\phi K_D + \left(1 - \frac{(1 + \phi) K_D}{K_s}\right)K_f \right]\left(1 + \Delta + \phi \frac{\partial \Delta}{\partial \phi}\right)}{\phi^2 (1 + \Delta)^2},\\
\frac{\partial G}{\partial \phi} &= \frac{\partial G_D}{\partial \phi},\\
\frac{\partial \rho}{\partial \phi} &= \rho_f - \rho_s,
\end{split}
\end{aligned}
\end{equation}
where
\begin{equation}
\begin{aligned}
\begin{split}
\frac{\partial K_D}{\partial \phi} &= -K_s\frac{1+ cs}{(1 + cs \phi)^2},\\
\frac{\partial \Delta}{\partial \phi} &= -\frac{K_f}{K_s\phi^2}\left(1 - \frac{1}{1 + cs \phi}\right) + \frac{K_f (1 - \phi)}{K_s\phi}\frac{cs}{(1+cs\phi)^2},\\
\frac{\partial G_D}{\partial \phi} &= -G_s\frac{1+ \frac{3}{2} cs}{(1 + \frac{3}{2} cs \phi)^2}.
\end{split}
\end{aligned}
\end{equation}

With respect to $C$, we have 
\begin{equation}
\begin{aligned}
\begin{split}
\frac{\partial K}{\partial C} &= \frac{ \phi \frac{\partial K_D}{\partial C} + \frac{K_f (1 + \phi)}{K_s^2} \left( \frac{\partial K_s}{\partial C}  K_D -\frac{\partial K_D}{\partial C} K_s \right)}{\phi (1 + \Delta)} -\\
& \frac{\left[\phi K_D + \left(1 - \frac{(1 + \phi) K_D}{K_s}\right)K_f \right] \phi \frac{\partial \Delta}{\partial C}}{\phi^2 (1 + \Delta)^2},\\
\frac{\partial G}{\partial C} &= \frac{\partial G_D}{\partial C},\\
\frac{\partial \rho}{\partial C} &= (1-\phi)\frac{\partial \rho_s}{\partial C},
\end{split}
\end{aligned}
\end{equation}
where
\begin{equation}
\begin{aligned}
\begin{split}
\frac{\partial \Delta}{\partial C} &= \frac{(\phi - 1) K_f}{\phi K_s^2}\left(1 - \frac{1}{1 + cs \phi}\right) \frac{\partial K_s}{\partial C},\\
\frac{\partial K_s}{\partial C} &= K_c - K_q ,\qquad &&\frac{\partial G_s}{\partial C} = G_c - G_q,\\
\frac{\partial K_D}{\partial C} &= \frac{1-\phi}{1+cs\phi}\frac{\partial K_s}{\partial C},  \qquad &&\frac{\partial G_D}{\partial C} = \frac{1 - \phi}{1 + \frac{3}{2} cs \phi} \frac{\partial G_s}{\partial C},\\
\frac{\partial \rho_s}{\partial C} &= (\rho_c - \rho_q).
\end{split}
\end{aligned}
\end{equation}

Finally, with respect to $S_w$, we have 
\begin{equation}
\begin{aligned}
\begin{split}
\frac{\partial K}{\partial S_w} &= \frac{ \left( 1- \frac{(1+ \phi) K_D}{K_s}  \right)\frac{\partial K_f}{\partial S_w}}{\phi (1 + \Delta)} -\\
& \frac{\left[\phi K_D + \left(1 - \frac{(1 + \phi) K_D}{K_s}\right)K_f \right] \phi \frac{\partial \Delta}{\partial S_w}}{\phi^2 (1 + \Delta)^2},\\
\frac{\partial G}{\partial S_w} &=  0,\\
\frac{\partial \rho}{\partial S_w} &= \phi(\rho_w - \rho_h),
\end{split}
\end{aligned}
\end{equation}
where
\begin{equation}
\begin{aligned}
\begin{split}
\frac{\partial K_f}{\partial S_w} &= K_w - K_h, \\
\frac{\partial \Delta}{\partial S_w} &= \frac{\Delta }{K_f}\frac{\partial K_f}{\partial S_w}.
\end{split}
\end{aligned}
\end{equation}
Gathering the previous terms, the gradient of the cost function with respect to model parameters in PCS parameterization can be calculated as

\begin{equation}
\left[\begin{array}{c}
    \frac{\partial \chi}{\partial \phi} \\
    \frac{\partial \chi}{\partial C}\\
    \frac{\partial \chi}{\partial S_w}
\end{array}\right]=
\left[\begin{array}{c c c}
    \frac{\partial V_P}{\partial \phi} & \frac{\partial V_S}{\partial \phi}& \frac{\partial \rho}{\partial \phi}\\
  \frac{\partial V_P}{\partial C} & \frac{\partial V_S}{\partial C} & \frac{\partial \rho}{\partial C} \\
\frac{\partial V_P}{\partial S_w} & \frac{\partial V_S}{\partial S_w}  & \frac{\partial \rho}{\partial S_w}
\end{array} \right]
\left[\begin{array}{c}
    \frac{\partial \chi}{\partial V_P} \\
    \frac{\partial \chi}{\partial V_S}\\
    \frac{\partial \chi}{\partial \rho}
\end{array}\right].
\label{eq:grad_vd2pcs}
\end{equation}

\end{appendices}

\end{document}